# The Novel Coronavirus, 2019-nCoV, is Highly Contagious and More Infectious Than Initially Estimated

**Authors:** Steven Sanche[1,2,†], Yen Ting Lin[3,†], Chonggang Xu[4], Ethan Romero-Severson[1], Nicolas W. Hengartner[1], Ruian Ke[1,*]

**Affiliations:**
[1]T-6 Theoretical Biology and Biophysics, Theoretical Division, Los Alamos National Laboratory, NM87544, USA.
[2]T-CNLS Center for Nonlinear Studies, Los Alamos National Laboratory, NM87544, USA.
[3]CCS-3 Information Sciences Group, Computer, Computational and Statistical Sciences Division, Los Alamos National Laboratory, Los Alamos, New Mexico 87545, USA
[4]EES-14 Earth Systems Observations Group, Earth and Environmental Sciences Division, Los Alamos National Laboratory, Los Alamos, New Mexico 87545, USA

[†]S.S. and Y.T.L. contributed equally to the work.

[*]Correspondences should be addressed to:
Ruian Ke
Email: rke@lanl.gov
Phone: 1-505-667-7135
Mail: Mail Stop K710,
T-6 Theoretical Biology and Biophysics,
Los Alamos National Laboratory,
NM87544, USA.

**Short title:** The 2019 novel coronavirus is highly infectious

**Word counts:**
Abstract: 124
Main text including references and figure captions: 3,945

**Abstract**

The novel coronavirus (2019-nCoV) is a recently emerged human pathogen that has spread widely since January 2020. Initially, the basic reproductive number, $R_0$, was estimated to be 2.2 to 2.7. Here we provide a new estimate of this quantity. We collected extensive individual case reports and estimated key epidemiology parameters, including the incubation period. Integrating these estimates and high-resolution real-time human travel and infection data with mathematical models, we estimated that the number of infected individuals during early epidemic double every 2.4 days, and the $R_0$ value is likely to be between 4.7 and 6.6. We further show that quarantine and contact tracing of symptomatic individuals alone may not be effective and early, strong control measures are needed to stop transmission of the virus.

**One-sentence summary**

By collecting and analyzing spatiotemporal data, we estimated the transmission potential for 2019-nCoV.

**Main Text**

2019-nCoV is the etiological agent of the current rapidly growing outbreak originating from Wuhan, Hubei province, China (*1*). At the end of December 2019, 41 cases of 'pneumonia of unknown etiology' were reported by the Wuhan Municipal Health Committee (*2*). On January 1, 2020, the Huanan Seafood Wholesale Market in Wuhan, which was determined to be the epicenter of the outbreak, was closed. Seven days later, the causative agent of new disease was formally announced by China CDC as 2019-nCoV. Human-to-human transmission was later reported, i.e. infection of medical workers reported by the news and infection of individuals with no recent history of Wuhan visit (*3*). In response, China CDC upgraded the emergency response to Level 1 (the highest level) on January 15. By January 21, 2019-nCoV infection had spread to most of the other provinces. On January 23, the city of Wuhan was locked down/quarantined, all transportations into and out of the city and all mass gatherings was canceled. However, the number of confirmed cases has continued to increase exponentially since January 16 (Fig. 1A and B). On January 30, the World Health Organization (WHO) declared the outbreak a public health emergency of international concern (*4*) . As of February 5, 2020, the virus outbreak lead to more than 24,000 total confirmed cases and 494 deaths, and the virus has spread to 25 countries. Initial estimates of the growth rate of the outbreak based on early case count data in Wuhan and international flight data up to mid-January were 0.1 per day (a doubling time between 6-7 days) and a basic reproductive number, $R_0$ (defined as the average number of secondary cases an index case infects when it is introduced in a susceptible population), of 2.2 and 2.7 (*1, 5*); however, the rates of growth in the number of confirmed cases during late January (Fig. 1A and B) suggest a doubling time much shorter than 6-7 days.

Here, with more up-to-date and high-resolution datasets across China until the end of January, we estimated that the exponential growth rate and $R_0$ are much higher than these previous estimates. We improved on previous estimates in three distinct ways. First, we used an expanded dataset of individual case reports based on our collection and direct translations of documents published daily from official health commissions across provinces and special municipalities in China (see Data Collection in Supplementary Materials). Second, we integrated high-resolution real-time domestic travel data in China. Third, to address the issue of potential data collection and methodological bias or incomplete control of confounding variables, we implemented two distinct modeling approaches using different sets of data. These analyses produced estimates of the exponential growth rates that are consistent with one another and higher than previous estimates.

A unique feature of our case report dataset (Table S1) is that it includes case reports of many of the first or the first few individuals who were confirmed with the virus infection in each province, where dates of departure from Wuhan were reported. All together, we collected 140 individual case reports (Table S1). These reports include demographic information including age, sex and location of hospitalization, as well as epidemiological information including potential time periods of infection, dates of symptom onset, hospitalization and case confirmation.

Using this dataset, we estimated the basic parameter distributions of durations from initial exposure to symptom onset to hospitalization to discharge or death. Our estimate of the time from initial exposure to symptom onset is 4.2 days with a 95% confidence interval (CI for short below) between 3.5 and 5.1 days (Fig. 1C). This estimated period is about 1 day shorter and has lower variance than a previous estimate (*1*). The shorter time is likely caused by the expanded temporal range of our data that includes cases occurring after broad public awareness of the disease. Patients reported in the Li et al. study (*1*) are all from Wuhan and most had symptom onset before mid-January; in our dataset, many patients had symptom onset during or after mid-January and were reported in provinces other than Hubei province (where Wuhan is the capital). The time from symptom onset to hospitalization showed evidence of time dependence (Fig. 1D and S1). Before January 18, the time from symptom onset to hospitalization was 5.5 days (CI: 4.6 to 6.6 days); whereas after January 18, the duration shortened significantly to 1.5 days only (CI: 1.2 to 1.9 days) ($p$-value <0.001 by Mann-Whitney U test). The change in the distribution coincides with the period when infected cases were first confirmed in Thailand, news reports of potential human-to-human transmission and upgrading of emergency response level to Level 1 by China CDC. The emerging consensus about the risk of 2019-nCoV likely led to significant behavior change in symptomatic people seeking more timely medical care over this period. We also found that the time from initial hospital admittance to discharge is 11.5 days (Fig. 1E; CI: 8.0 to 17.3 days) and the time from initial hospital admittance to death is 11.2 days (Fig. 1F; CI: 8.7 to 14.9 days).

Moving from empirical estimates of basic epidemiological parameters to an understanding of the actual epidemiology of 2019-nCoV requires model-based inference. We thus used mathematical models to integrate the empirical estimates with spatiotemporal domestic travel and infection data outside of Hubei province to infer the outbreak dynamics in Wuhan. Inference based on data outside of Hubei is more reliable because, as a result of the awareness of the risk of virus transmission, other provinces implemented intensive surveillance system to detect individuals with high temperatures and closely track travelers out of Wuhan using digital data to identify infected individuals (*6*) as the outbreak in Wuhan unfolded.

We collected real-time travel data during the epidemic using the Baidu® Migration server (Fig. 2A and Table S2). The server an online platform summarizing mobile phone travel data through Baidu® Huiyan [https://huiyan.baidu.com/]. Baidu® Huiyan is a widely used positioning system in China. It processes >120 billion positioning requests daily through GPS, WIFI and other means [https://huiyan.baidu.com/]. Therefore, the data represents a reliable, real-time and high-resolution source of travel patterns in China. We extracted daily travel data from Wuhan to each of the provinces. We found that in general, between 40,000 to 140,000 people in Wuhan traveled to destinations outside of Hubei province daily before the lock-down of the city on January 23, with travel peaks on January 9, 21 and 22 (Fig. 2B). Thus, it is likely that this massive flow of people from Hubei province during January facilitated the rapid dissemination of virus.

We integrated the travel data into our inferential models using two approaches. The rationale of the first model, the 'first-arrival' approach, is that an increasing fraction of people infected in Wuhan increases the likelihood that one such case is exported to the other provinces. Hence, how soon new cases are observed in other provinces can inform disease progression in Wuhan (Fig. 2C). This has similarities with earlier analyses to estimate the size of the 2019-nCoV outbreak in Wuhan based on international travel data (*5, 7, 8*), though inference based infected cases outside of China may suffer large uncertainty due to the low volume of international travel. In our model, we assumed exponential growth for the infected population $I*$ in Wuhan, $I^*(t) = e^{r(t-t_0)}$, where $r$ is the exponential growth rate and $t_0$ is the time of the exponential growth initiation, i.e. $I^*(t_0) = 1$. Note that $t_0$ is likely to be later than the date of the first infection event, because multiple infections may be needed before the onset of exponential growth (*9*). We used travel data to each of the provinces (Table S3) and the earliest times that an infected individual arrived at a province across a total of 26 provinces (Fig. 2D) to infer $r$ and $t_0$ (see Supplementary Materials for details). Model predictions of arrival times in the 26 provinces fitted the actual data well (Fig. S2). We estimated that the date of the beginning of an exponential growth is December 20, 2019 (CI: December 11 to 26). This suggests that human infections in early December may be due to spillovers from the animal reservoir or limited chains of transmission (*10, 11*). The growth rate of the outbreak is estimated to be 0.29 per day (CI: 0.21 to 0.37 per day), a much higher rate than two recent estimates (*1, 5*). This growth rate corresponds to a doubling time of 2.4 days. We further estimated that the total infected population size in Wuhan was approximately 4,100 (CI: 2,423 to 6,178) on January 18, which is remarkably consistent with a recently posted estimate (*7*). The estimated number of infected individuals is 18,700 (CI: 7,147, 38,663) on January 23, i.e. the date when Wuhan started lock down. We projected that without any control measure, the infected population would be approximately 233,400 (CI: 38,757 to 778,278) by the end of January (Fig. S3).

An alternative model, the 'case count' approach, used daily case count data between January 19 and 26 from provinces outside of Hubei to infer the initiation and the growth rate of the outbreak. We restricted the data to this period because during this time infected persons found outside of Hubei province generally reported visiting Wuhan within 14 days of becoming symptomatic, i.e. cases during that time period were indicative of the dynamics in Wuhan. We developed a meta-population model based on the classical SEIR model (*12*). We assumed a deterministic exponential growth for the infected populations in Wuhan, whereas in other provinces, we represented the trajectory of infected individuals who travelled from Wuhan using a stochastic agent-based model. The transitions of the infected individuals from symptom onset to hospitalization and then to case confirmation were assumed to follow the distributions inferred from the case report data (see Supplementary Materials for detail). Simulation of the model using best fit parameters showed that the model described the observed case counts over time well (Fig. 2E). The estimated date of exponential growth initiation is December 16, 2019 (CI: December 12 to Dec 21) and the exponential growth rate is 0.30 per day (CI: 0.26 to 0.34 per day). These estimates are consistent with estimates in the 'first arrival' approach (Fig. 2F and G, and Fig. S4).

We note that in both approaches, we assumed perfect detection of infected cases outside of Hubei province, i.e. the dates of first arrival and the number of case counts are accurate. This could be a reasonable assumption to make for symptomatic individuals because of the intensive surveillance implemented in China, for example, tracking individuals' movement from digital transportation data (*6*). However, it is possible that a fraction of infected individuals, for example, individuals with mild or no symptoms (*13*), were not hospitalized, in which case we will underestimate the true size of the infected population in Wuhan. We undertook sensitivity analyses to investigate how our current estimates are affected by this issue using both approaches (see Supplementary Materials for detail). We found that if a proportion of cases remained undetected, the time of exponential initiation would be earlier than December 20, translating into a larger population of infected individuals in January, but the estimation of the growth rate remained the same. Overall, the convergence of the estimates of the exponential growth rate from the two approaches emphasizes the robustness of our estimates to model-dependent assumptions.

Our estimated outbreak growth rate is significantly higher than two recent reports where the growth rate was estimated to be 0.1 per day (*1, 5*). This estimate were based on early case counts from Wuhan (*1*) or international air travel data (*5*). However, these data suffer from important limitations. The reported case counts in Wuhan during early outbreak are likely to be underreported because of many factors, and because of the low numbers of individuals traveling abroad compared to the total population size in Wuhan, inference of the infected population size and outbreak growth rate from infected cases outside of China suffers from large uncertainty (*7, 8*). Our estimated exponential growth rate, 0.29/day (a doubling time of 2.4 days) is consistent the rapidly growing outbreak during late January (Fig. 1A).

Using the exponential growth rate, we next estimated the range of the basic reproductive number, $R_0$. It has been shown that this estimation depends on the distributions of the latent period (defined as the period between the times when an individual infected and become infectious) and the infectious period (*14*). For both periods, we assumed a gamma distribution and varied the mean and the shape parameter of the gamma distributions in a large range to reflect the uncertainties in these distributions (see Supplementary Materials). It is not clear when an individual becomes infectious; thus, we considered two scenarios: 1) the latent period is the same as the incubation period, and 2) the latent period is 2 days shorter than the incubation period, i.e. individuals start to transmit 2 days before symptom onset. Integrating uncertainties in the exponential growth rate estimated from the 'first arrival' approach and the uncertainties in the duration of latent and infectious periods, we estimated the values of $R_0$ to be 6.3 (CI: 3.3 to 11.3) and 4.7 (CI: 2.8 to 7.6), for the first and second scenarios, respectively (Fig. 3A). When using the estimates from the 'case count' approach, we estimated slightly higher $R_0$ values of 6.6 (CI: 4.0 to 10.5) and 4.9 (CI: 3.3 to 7.2), for the first and second scenarios, respectively (Fig. S5). Overall, we report $R_0$ values are likely be between 4.7 and 6.6 with a CI between 2.8 to 11.3. We argue that the high $R_0$ and a relatively short incubation period lead to the extremely rapid growth of the of 2019-nCoV outbreak as compared to the 2003 SARS epidemic where $R_0$ was estimated to be between 2.2 to 3.6 (*15, 16*).

The high $R_0$ values we estimated have important implications for disease control. For example, basic theory predicts that the force of infection has to be reduced by $1-\frac{1}{R_0}$ to guarantee extinction of the disease. At $R_0 = 2.2$ this fraction is only 55%, but at $R_0 = 6.7$ this fraction rises to 85%. To translate this into meaningful predictions, we use the framework proposed by Lipsitch et al (*16*) with the parameters we estimated for 2019-nCoV. Importantly, given the recent report of transmission of the virus from asymptomatic individuals (*13*), we considered the existence of a fraction of infected individuals who is asymptomatic and can transmit the virus (see Supplementary Materials). Results show that if as low as 20% of infected persons are asymptomatic and can transmit the virus, then even 95% quarantine efficacy will not be able to contain the virus (Fig. 3B). Given the rapid rate of spread, the sensitivity of control effort effectiveness to asymptomatic infections and the potential of transmission before symptom onset, we need to be aware of the difficulty of controlling 2019-nCoV once it establishes in a new population (*17*). Future field, laboratory and modeling studies aimed to address the unknowns, such as the fraction of asymptomatic individuals, the time when individuals become infectious and the existence of superspreaders are needed to accurately predict the impact of various control strategies (*9, 17*).

Fortunately, we see evidence that control efforts have a measurable effect on the rate of spread. Since January 23, Wuhan and other cities in Hubei province implemented vigorous control measures, such as closing down transportation and mass gatherings in the city; whereas, other provinces also escalated the public health alert level and implemented strong control measures. We noted that the growth rate of the daily number of new cases in provinces outside of Hubei slowed down gradually since late January (Fig. 3B). Due to the closure of Wuhan (and other cities in Hubei), the number of cases reported in other provinces during this period shall start to track local infection dynamics rather than imports from Wuhan. We estimated that the exponential growth rate is decreased to 0.14 per day (CI: 0.12 to 0.15 per day) since January 30. Based on this growth rate and an $R_0$ between 4.7 to 6.6 before the control measures, a calculation following the formula in Ref. (*14*) suggested that a growth rate decreasing from 0.29 per day to 0.14 per day translates to a 50%-59% decrease in $R_0$ to between 2.3 to 3.0. This is in agreement with previous estimates of the impact of effective social distancing during 1918 influenza pandemic (*18*). Thus, the reduction in growth rate may reflect the impact of vigorous control measures implemented and individual behavior changes in China during the course of the outbreak.

The 2019-nCoV epidemic is still rapidly growing and spread to more than 20 countries as of February 5, 2020. Here, we estimated the growth rate of the early outbreak in Wuhan to be 0.29 per day (a doubling time of 2.4 days), and the reproductive number, $R_0$, to be between 4.7 to 6.6 (CI: 2.8 to 11.3). Among many factors, the Lunar New Year Travel rush in early and mid-January 2020 may or may not play a role in the high outbreak growth rate, although SARS epidemic also overlapped with the Lunar New Year Travel rush. How contiguous the 2019-nCoV is in other countries remains to be seen. If the value of $R_0$ is as high in other countries, our results suggest that active and strong population-wide social distancing efforts, such as closing down transportation system, schools, discouraging travel, etc., might be needed to reduce the overall contacts to contain the spread of the virus.

**Acknowledgments.** We would like to thank Alan Perelson and Christiaan van Dorp for suggestions and critical reading of the manuscript and Weili Yin for help with collecting and translating documents from provincial health commission websites. **Funding:** SS and RK would like to acknowledge funding from DARPA (HR0011938513). CX acknowledges the support from the Laboratory Directed Research and Development (LDRD) Program at Los Alamos National Laboratory. The funders had no role in study design, data collection and analysis, decision to publish, or preparation of the manuscript. **Author contributions:** RK and NH conceived the project; RK collected data; SS, YTL, CX and RK performed analyses; SS, YTL, ERS, NH and RK wrote and edited the manuscript. **Competing interests:** authors declare no competing interests. **Data and materials availability:** all data is available in the main text or the supplementary materials.

**Supplementary Materials:**

Supplementary Text

Figures S1-S6

Tables S1-S3

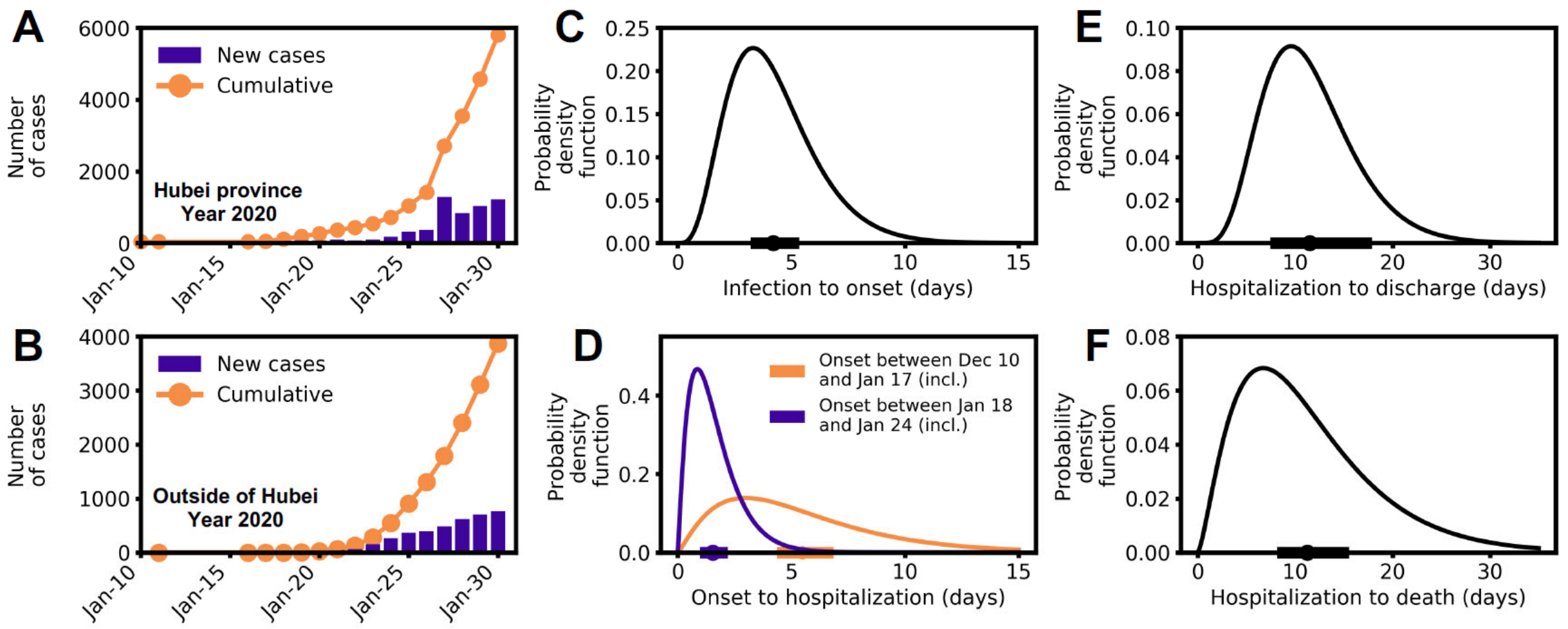

**Fig. 1.** Epidemiological characteristics of early dynamics of 2019-nCoV outbreak in China. **(A-B)** Daily new and cumulative confirmed cases in Hubei province (A) and provinces other than Hubei in China (B). **(C-F)** Distributions of key epidemiological parameters, including the durations between infection and symptom onset (C), between symptom onset to hospitalization (D), between hospitalization to discharge (E) and between hospitalization to death (F). Filled circles and bars on x-axes denote the estimated mean and 95% confidence intervals.

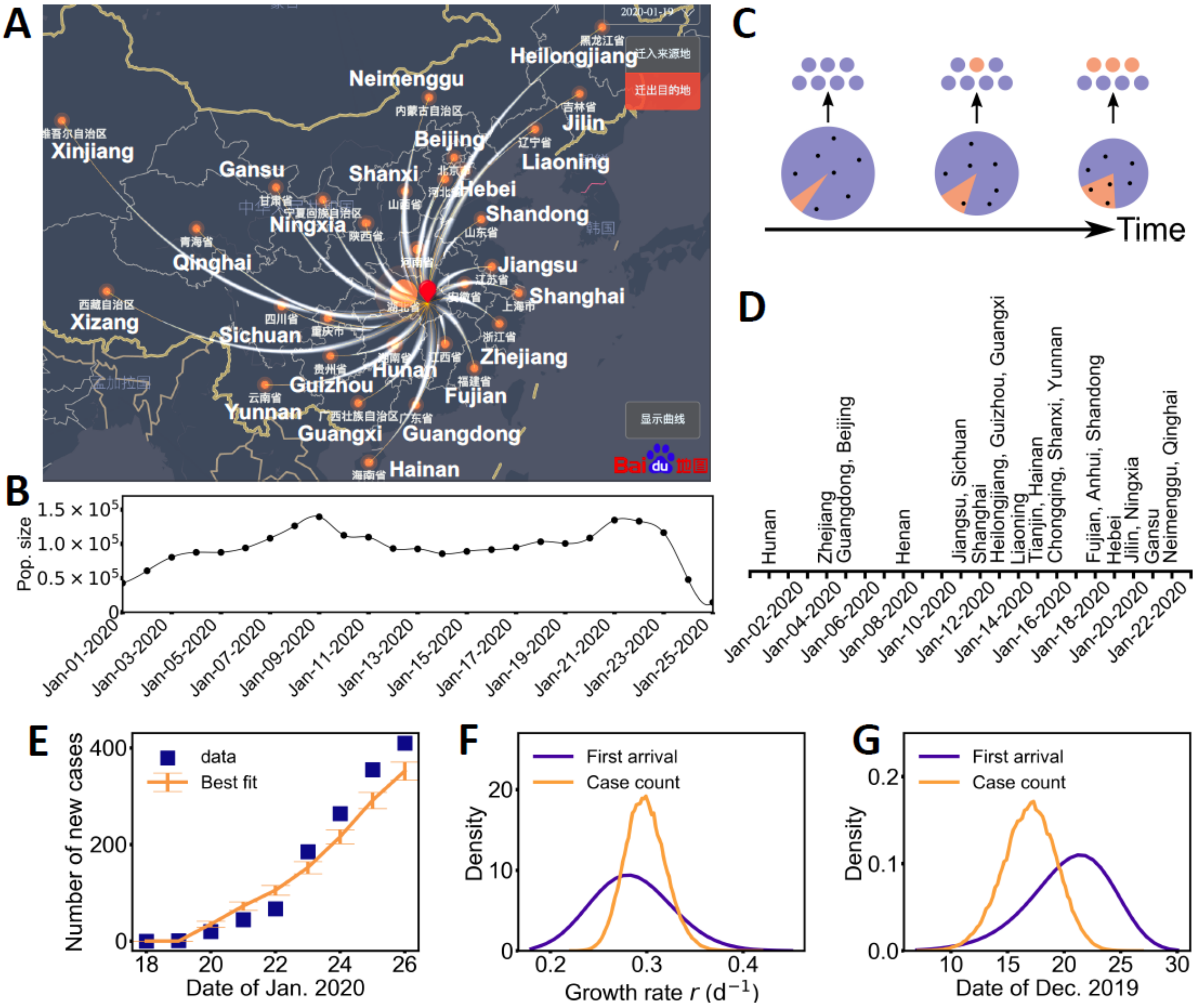

**Fig. 2.** Two different approaches using high-resolution travel data reached consistent estimates of the exponential growth rate and the date of exponential growth initiation of the 2019-nCoV outbreak. **(A)** A modified snapshot of the Baidu® Migration online server interface showing the migration pattern out of Wuhan (red dot) on January 19, 2020. Thickness of curved white lines denotes the size of the traveler population to each province. The names of most of the provinces are shown in white. **(B)** Estimated daily population sizes of travelers from Wuhan, Hubei province to other provinces. **(C)** A schematic illustrating the export of infected individuals from Wuhan. Travelers (dots) are assumed to be random samples from the total population (the whole pie). Because of the growth of the infected population (orange pie) and the shrinking size of the total population in Wuhan over time, it is more and more likely infected individuals travel to other provinces (orange dots). **(D)** The dates of documented first arrivals of infected cases in 26 provinces. Names of provinces were shown vertically. **(E)** Best fit of the 'case count' model to daily counts of new cases (including only imported cases) in provinces other than Hubei. The standard deviations of the sample distribution are shown as the error bars. **(F and G)** The marginalized likelihoods of the growth rate $r$ (F) the exponential growth initiation time (G) are consistent between the 'first arrival' model and the 'case count' model.

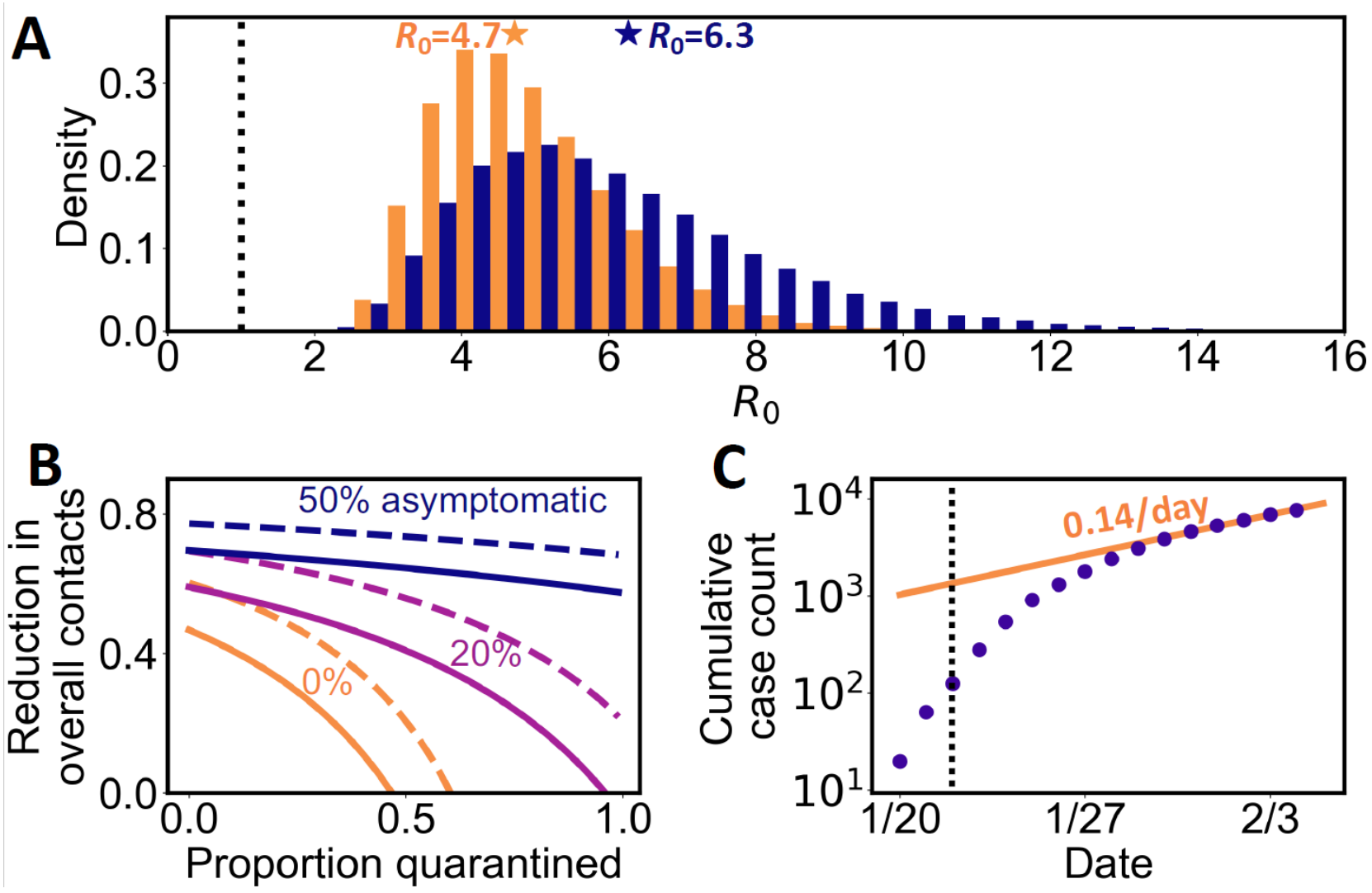

**Fig. 3.** Estimation of the basic reproductive number, $R_0$, and the impact of control measures. **(A)** Histograms and the means (stars) of estimated $R_0$ assuming individuals become infectious at symptom onset (blue) or 2 days before symptom onset (orange). The dotted line denotes $R_0$=1. **(B)** The levels of minimum efforts (lines) of intervention strategies needed to control the virus spread. Strategies considered are quarantine of symptomatic individuals and individuals who had contacts with them (x-axis) and population-level efforts to reduce overall contact rates (y-axis). Different colored lines denote different assumptions of the fraction of asymptomatic individuals in the infected population. Solid and dashed lines correspond to $R_0$=4.7 and 6.3 (i.e. the estimated means of $R_0$), respectively. **(C)** The cumulative number of cases outside of Hubei province in late January 2020. The growth rate decreased to 0.14 per day since January 30. The dashed black line shows January 23 when Wuhan is locked down.

# Supplementary Materials for

# "The Novel Coronavirus, 2019-nCoV, is Highly Contagious and More Infectious Than Initially Estimated"

Steven Sanche[1,2,†], Yen Ting Lin[3,†], Chonggang Xu[4], Ethan Romero-Severson[1], Nicolas W. Hengartner[1], Ruian Ke[1,*]

[†]S.S. and Y.T.L. contributed equally to the work.

[*]Correspondences to: Ruian Ke (rke@lanl.gov)

**This PDF file includes:**

Supplementary Text
Figs. S1 to S5
Tables S1 to S3

# Supplementary Text

## Data Collection

*Case count and individual case reports*

We collected and translated reports from documents published daily from the China CDC website and official websites of health commissions across provinces and special municipalities in China (website URLs are available upon request). We collected daily counts of confirmed cases in each province as well as 140 individual case reports (Table S1). Many of the individual reports were also published on the China CDC official website (http://www.chinacdc.cn/jkzt/crb/zl/szkb_11803/) and the China CDC weekly bulletin (in English) (http://weekly.chinacdc.cn/news/TrackingtheEpidemic.htm). Our dataset includes demographic information including age, sex as well as epidemiological information including dates of symptom onset, hospitalization, case confirmation, discharge or death. Most of the health commissions in provinces and special municipalities documented and published detailed information of the first or the first few cases confirmed with 2019-nCoV infection. As a result, this dataset includes case reports of many of the first or the first few individuals who were confirmed with the virus infection in each province, where dates of departure from Wuhan were available.

*Travel data*

We used the Baidu® Migration server (https://qianxi.baidu.com/) to estimate the number of daily travelers in and out Wuhan (Table S2). Specifically, we extracted from the server the Immigration Index and Emigration Index for Wuhan, which are linearly related to the number of travelers going in and out of Wuhan, respectively, based on cell phone positioning data. We also extracted the fraction of individuals who went to or came from a particular province. It has been reported that there were 5 million people going out of Wuhan between the start of the Chinese New Year travel rush and January 23 (https://www.washingtonpost.com/world/asia_pacific/china-coronavirus-live-updates/2020/01/30/1da6ea52-4302-11ea-b5fc-eefa848cde99_story.html; accessed Feb. 2, 2020). This allowed us to calibrating the Emigration Index and estimated the number of daily travelers to or from a particular province, and thus the fraction of people traveling to or from a particular province (Table S3). These data were used in mathematical models to estimate the s

## Estimating distributions of epidemiological parameters from individual reports

We used the first confirmed cases in provinces other than Hubei to inform the time between patient infection and the onset of symptoms ($n = 24$). These individuals had all traveled to Wuhan a short time preceding symptoms onset. Since these individuals were the first cases detected in the province, it is likely that the infection occurred during their recent stay in Wuhan. We approximated the time of infection as the middle time point of their stay. Because the delays between infection and symptoms onset vary between patients, we modeled the delay using a gamma distribution, as its support is nonnegative and it permits relatively large delays as compared to the median. Figure 1 in the main text presents results from fitting the distribution to the data.

The fitting procedure was performed by maximizing the likelihood of observed delays between infection and symptoms onset. For a single observation, the individual likelihood is the gamma density function evaluated at the infection-to-onset delay. Some of the delays were censored, i.e. bounded by a certain value. For example, in some cases, only the times of infection and hospitalization were reported, and the time of symptom onset was missing in the case report. In such cases, we assumed that the missing onset time is bounded between times of infection and hospitalization. Then, the likelihood for this observation is equal to the cumulative gamma distribution evaluated at this censored value, i.e., the time when the patient was hospitalized. The maximum likelihood estimates (MLEs) are the shape and scale parameters that maximize the sum over all observations of the individual log-likelihoods. We used differential_evolution in scipy.optimize library (Python) to perform maximization. A stochastic algorithm was implemented in the optimization procedure to avoid being trapped in local minima.(1) The likelihood-based confidence intervals was computed by methods reported in Raue et al.(2)

A similar approach was adopted to fit distributions to the time between symptom onset and hospitalization ( $n = 96$ ), between hospitalization and discharge ( $n = 6$ ), and between hospitalization and death ($n = 23$). The reported dates for these events was obtained directly from official sources. Data from cases originating from all over China and neighboring countries were used for distribution fitting. Detailed patient-level data is provided in Table S1.

## The ‘first-arrival’ model: Inferring disease dynamics in Wuhan using the first-arrival times at other provinces

In this model, we used the first-arrival time of a patient who traveled from Wuhan to a specific other province and was later confirmed to have been infected by the 2019-nCoV. The rationale behind our approach is that an increasing fraction of people infected in Wuhan increases the likelihood that one such case is exported to the other provinces. Hence, how soon new cases are observed in other provinces can inform the disease progression in Wuhan. We hypothesize that this information is more reliable because the infected population in Wuhan needs to sufficiently large to allow probable export of one infected individual. The flow of expected cases depends on the flow of travelers to each province and on the proportion of the Wuhan population that is infected by the virus.

We first estimated the daily number of travelers from Wuhan to each of the China provinces. For this purpose, we used Wuhan’s daily migration index to other provinces and the daily distribution of traveler destinations from Wuhan (see Data Collection). When assuming linearity between the migration index and the total number of exported individuals, it can be estimated that a migration index of 1 is approximately equal to 5 million individuals over the sum of migration indexes from January 10 to January 25, 2020 (it was reported that 5 million individuals left Wuhan during that period; see Data Collection section). The total number of daily Wuhan travelers to a province at a certain date was then set equal to the number of travelers estimated from the migration index times the fraction of the population having traveled to this province. Results from estimation are reported in Table S2.

An infected traveler may be pre-symptomatic, i.e. this individual may have been exposed to the virus ($E$) and not have developed symptoms or be already symptomatic ($I$). In fact, for many individuals, infection onset was recorded days after the time of their departure from Wuhan (see Table S1). Assuming travelers represent a random sample of the whole population, it follows that the probability that a traveler is infected is equal to the number of exposed or infected individuals in Wuhan ($I^* = E + I$) over the total Wuhan population ($N(t)$). The total population size varied during the infection period. We estimated the population size by using the daily inflow and outflow of individuals from Wuhan (see Table S2). In order to represent the beginning of an outbreak, we modeled an exponential increase in the size of exposed and infected population over time $t$:

$$I^*(t) = e^{r(t-t_0)} \tag{1}$$

where $r$ is the infection growth rate and $t_0$ is the time of onset of exponential outbreak.

Equation (1) allows a simple analytic expression of the likelihood of arrival times for the first cases in each of the provinces other than Hubei. For a specific province, indexed by $i$, we modeled the arrival of new cases in each province during short time intervals as a Poisson random process $X_t^{(i)}$. Note that the rate parameter of this Poisson process, $\lambda(t) = I^*(t)\, \kappa_i(t)/N(t)$ depends on the time-varying sum of exposed and symptomatic populations $I^*(t)$, the time varying flow of population $\kappa_i(t)$ transported from Wuhan to the province $i$ and the time varying population size. It can be shown mathematically (3) that the probability that no exposed or symptomatic traveler arrived to province $i$ during a short time interval $(t, t + \Delta t)$ , $\Delta t \ll 1$ is:

$$\mathbb{P}\left\{X_{t+\Delta t}^{(i)} - X_t^{(i)} = 0\right\} \approx \exp\left(-\frac{I^*(t)\kappa_i(t)}{N(t)}\Delta t\right). \tag{2}$$

We assume no delay was incurred due transportation in our model. Equation (222) is valid for any $t > 0$, and because the overall process is Markovian, we can formulate the probability that the time of arrival of the first case in province $i$, $T^{(i)}$, is later than $t$ by:

$$\mathbb{P}\left\{T^{(i)} > t\right\} = \lim_{\Delta t \to 0} \prod_{j=1}^{M} \mathbb{P}\left\{X_{j\Delta t}^{(i)} - X_{(j-1)\Delta t}^{(i)} = 0\right\} = \exp\left(-\int_{t_0}^{t} \frac{I^*(s)\kappa_i(s)}{N(s)} ds\right). \tag{3}$$

where $[t_0, t)$ was partitioned into $M$ equal intervals of $\Delta t = j(t - t_0)/M$, and we convert the Riemannian sum into an integral in the limit of $M \to \infty$ $(\Delta t \to 0)$. Finally, we apply $\mathrm{d}/\mathrm{d}t$ to $1 - \mathbb{P}\left\{T^{(i)} > t\right\}$ to obtain the probability density function (PDF) of the first-arrival time of province $i$:

$$\mathrm{PDF}_i(t) = \frac{I^*(t)\kappa_i(t)}{N(t)} \exp\left(-\int_{t_0}^{t} \frac{I^*(s)\kappa_i(s)}{N(s)} ds\right). \tag{4}$$

The form of the probability density function Eq. (4) was used to estimate the likelihood of observed arrival times in each province as a function of the growth rate $r$ and outbreak initiation time $t_0$. This likelihood was maximized, again using differential_evolution in scipy.optimize,(1) and the confidence intervals for $r$ and $t_0$ were obtained through profile likelihood.(2) Numerical integration was performed by discretizing time in daily time intervals, since both the flow of travelers and the population size in Wuhan were estimated daily.

*Sensitivity analyses for the 'first-arrival' model*

The arrival times were fitted using three versions of the above model. Each version made a different assumption on the probability that an infected or exposed individual having arrived at a location be later diagnosed with coronavirus. In the first sensitivity analysis, we assumed that this probability was 50%. In the second analysis, we assumed this probability to be 10%. Finally, we tested the assumption that this probability was 0% for cases having arrived before Dec 31st , 2019, after which point new infected arrivals had a 50% probability of being later diagnosed.

The model formulation above needed a small modification to perform analyses. The event $Y$: "no new arrival before time $t$ is later diagnosed with the infection" is now equivalent to "no arrival of an infected individual before time $t$", "one infected arrival before time $t$ remained undiagnosed", "two infected arrivals before time $t$ remained undiagnosed", etc. For a Poisson process with fixed parameter $\lambda$, the probability of $Y$ can be expressed as:

$$\mathbb{P}(Y) = \mathrm{e}^{-\lambda} + \sum_{k=1}^{\infty} \frac{(1-p)^k \lambda^k e^{-\lambda}}{k!} = e^{-\lambda p} \qquad (5)$$

where $p$ is the probability of detection. It follows that the modified PDF formulation for sensitivity analyses is:

$$\mathrm{PDF}_i(t) = \frac{I^*(t)\kappa_i(t)\, p}{N(t)} \exp\left(-\int_{t_0}^{t} \frac{I^*(s)\kappa_i(s)\, p}{N(s)} ds\right). \qquad (6)$$

This PDF was used instead of equation (4) to obtain maximum likelihood estimates of the growth rate and outbreak initiation date for sensitivity analyses.

*Results from sensitivity analyses*

The following are the maximum likelihood estimates for the growth rate and date of outbreak initiation in the hypothetical situations mentioned above. When the probability of detection of a case was set to 50%, the estimated growth rate was 0.29/day, while the time of outbreak initiation was Dec 18, 2019. The same estimates were obtained if we assumed no case could be detected for individuals having arrived from Wuhan before Dec 31, 2019. When the probability of detection of a case was set to 10%, the estimated growth rate remained 0.29/day, but the estimated outbreak initiation date was Dec 12, 2019.

**The 'case count' model: The SEIR-type hybrid stochastic model**

Model 1 fitted the time of arrival of the first confirmed case of each province. We used a different approach and a different dataset to infer disease dynamics. In particular, we constructed a hybrid stochastic model for inferring the disease dynamics using all confirmed cases outside Hubei. Since the measurements in Wuhan, Hubei may have been biased in early outbreak, it is our aim to use data from outside Hubei for the inference of the growth rate $r$ and the onset time $\tau$ (define $t = 0$ as 0:00 am, 1/1/2020), defined as the time when the sum of exposed and symptomatic populations $\approx 1$ in Wuhan. The model is hybrid in the sense that we will couple a deterministic and exponential growth to describe the outbreak in Wuhan and an agent-based model which describes the discrete population dynamics of the patients after they left Hubei to other provinces. We present a schematic diagram of the hybrid meta-population model in Supplementary Fig. 6 below.

*Deterministic and exponential dynamics in Wuhan*

We assume an exponential growth of the number of exposed ($E_W$, $W$ for *W*uhan) and symptomatic ($I_W$) populations in Wuhan over time, $E_W(t) = E_W(0)e^{rt}$ and $I_W(t) = I_W(0)e^{rt}$ from the onset. The overall growth rate $r$ is dominated by the largest eigenvalue of a sequential compound process, and given an $r$ value, the ratio $\phi \coloneqq E(0)/I(0)$ is asymptotic constant (4). Thus, given a growth rate parameter $r$ and an initial condition $E(t_0) + I(t_0) = 1$, we numerically compute the exposed population $E(t) = \phi(r)\left(1 + \phi(r)\right)^{-1} \exp\left(r(t - t_0)\right)$ and the symptomatic population $I(t) = \left(1 + \phi(r)\right)^{-1} \exp\left(r(t - t_0)\right)$.

*Agent-based model for patients who have left Wuhan to other provinces*

We assume that between 1/1 and 1/26, the populations in Wuhan are large and the dynamics can be reasonably approximate by the above deterministic and exponentially growing curves. However, the initial propagation of the disease to other provinces in China involves only a small population of exposed ($E_O$, $O$ for *O*thers) or symptomatic individuals who left Hubei province. In addition, the transitions between different phases of these patients, from exposed ($E_O$) to symptomatic ($I_O$), over to hospitalized ($H_O$), and finally to be confirmed by laboratory examinations ($C_O$) in other provinces are also variable (as we quantified in Fig. 1C-F). Consequently, the resulting population dynamics in other provinces is highly stochastic. We thus adopt an agent-based modeling approach and rely on kinetic Monte Carlo Sampling techniques detailed below to simulate the population dynamics in other provinces. With this approach, we aim to generate samples of (1) each individual patient who left Wuhan at a specific date, and (2) the individual's health status as the time progresses (susceptible, exposed, or symptomatic). The goal is to accumulate a large amount of Monte Carlo samples, by which we can compute the key summary statistics, i.e., the average case reported on each day between 1/18 and 1/26, to be compared against to the data. We achieve this by the following algorithmic procedures.

1. *Generate random number of infected populations leaving Wuhan*. We collected migration index which quantifies the fraction of total populations (14 million) in Wuhan that traveled to other provinces on each date $t_i = 1, \dots, 26$ (see Table S3). Assuming independence of an individual's health state (susceptible, exposed, or symptomatic) and the individual's migration decision (leaving to other provinces or not), on each date $t_i$, the exposed and symptomatic populations leaving Hubei can be modeled by two Bernoulli distributions, $B_E = \mathrm{Bernoulli}\left(E_W(t_i), \mu(t_i)\right)$ and $B_I = \mathrm{Bernoulli}\left(I_W(t_i), \mu(t_i)\right)$. Here, $E_W(t)$ and $I_W(t)$ are the exposed and symptomatic population in Wuhan, and are assigned to the nearest integers to the previously prescribed exponential growth, given model parameters $(r, t_0)$. Thus, to generate one stochastic sample path (realization), we generate Bernoulli-distributed random populations leaving Hubei on each day between 1/1 and 1/26 (both included), and model each of these *in silico* patients' health states by the following procedures.

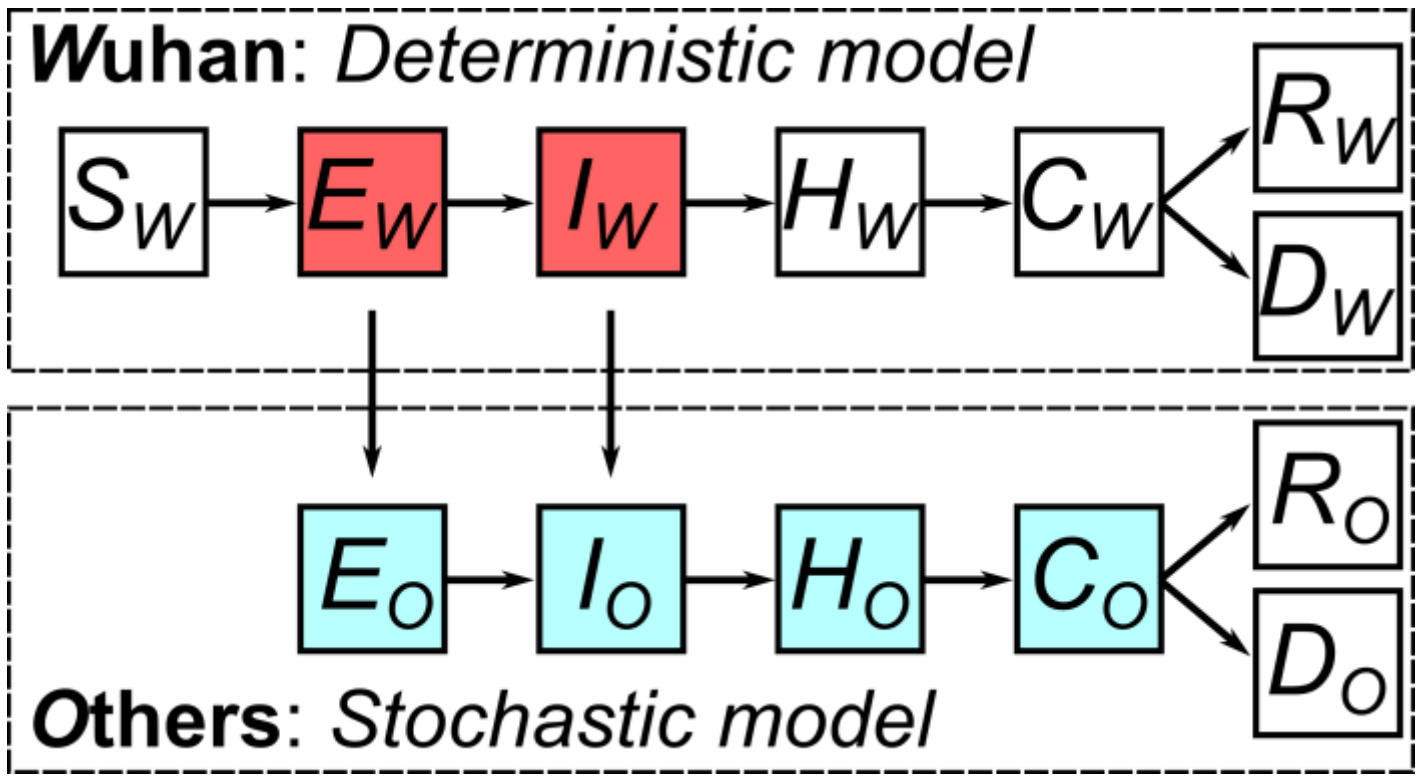

**Fig. S6. Schematic diagram of the proposed meta-population model.** Schematic diagram of the hybrid stochastic model. The model is a variant of the SEIR model with two geographic compartment, Wuhan (subscripted $W$) and other provinces (subscripted $O$). In Wuhan, a susceptible patient in compartment $S_W$ is first exposed and progresses to an exposed state ($E_W$), progressed to be infected ($I_W$), hospitalized ($H_W$), and then became a confirmed case ($C_W$), and either recovered ($R_W$) or deceased ($D_W$). A portion of ill population ($E_W$ and $I_W$) moved to other provinces and followed a similar progression. Because these populations are small and thus the dynamics are stochastic, we adopt an agent based approach to simulate the disease dynamics ( $E_O(t)$ , $I_O(t)$ , $H_O(t)$ and $C_O(t)$ ) in other provinces. The case reports on each day in other provinces were compared against the model's output, $C_O(t)$ to constrain the unknown initial onset and growth rate in Wuhan.

*2.* Generate the progression of the health state for each patient: We assume that each hypothetical patient generated by the above procedure would stochastically, identically and independently progress toward to be confirmed ($C_O$) and reported in one of the other provinces. If an individual was exposed ($E_O$) when s/he left Hubei at $t_i$, we generate a Gamma distributed random time $\Delta t_{E\to I} \sim \Gamma(\alpha_1, \beta_1)$ and update the individual's health state to symptomatic ($I_O$) at time $t_i + \Delta t_{E\to I}$ . We chose a time-dependent waiting-time distribution for the progression from symptomatic sate $I_O$ to reflect the two regimes we observed from the data (see main text): If $t_i + \Delta t_{E\to I}$ is before 1/18 (included), we generate a Gamma distributed random time $\Delta t_{I\to H} \sim \Gamma\left(\alpha_{2,1}, \beta_{2,1}\right)$ to model the waiting time for an infected patient to be hospitalized (otherwise, if it is later than 1/18, $\Delta t_{I\to H} \sim \Gamma\left(\alpha_{2,2}, \beta_{2,2}\right)$ ). Consequently, the patient's state is changed to $H_O$ at time $t_i + \Delta t_{E\to I} + \Delta t_{I\to H}$ . If $t_i + \Delta t_{E\to I} + \Delta t_{I\to H}$ is before 1/19, the patient would wait in the "H" state until 1/19 when the policy of case confirmation was announced and institutionalized. Then, the confirmation process is modelled by another Gamma distributed random time $\Delta t_{H\to C} \sim \Gamma(\alpha_3, \beta_3)$. The patient is then confirmed and reported at time $t_i + \Delta t_{E\to I} + \Delta t_{I\to H} + \Delta \mathrm{t}_{H\to C}$, and we add one more case report at the next integer (date of January). Similar procedure applied to a patient who had already progressed to the $I_W$ state before s/he left Hubei on date $t_i$, with the exception that the first random waiting time is neglected—the patient's confirmation time would be $t_i + \Delta t_{I\to H} + \Delta_{H\to C}$. We repeat the procedure for each *in-silico* patient who left

Wuhan between 1/1 and 1/26 (both included), and register the time when these patients were reported between 1/18 and 1/26 (both included).

*Parameter estimation and uncertainty quantification of* $(r, t_0)$

It is our task to infer the unknown parameters, exponential growth rate $r$ and exponential growth onset time $t_0$ by the number of confirmed cases reported between 1/18 and 1/26. This is possible because the information of the unknown parameters $(r, t_0)$ have an impact of the deterministic growths of the exposed $E_W(t)$ and symptomatic population $I_W(t)$, which in turn have an impact on the random populations which have left Hubei on each date. These populations follow statistically quantified processes until the final confirmation outside of Hubei, and can be compared against the reported data.

An error measure is devised to assess the quality of fit of the model given a set of parameters $(r, t_0)$ by the following procedures. For each parameter set, we generate $2^{13} = 8192$ Monte Carlo samples. On each date $t_i$, the $j^{\text{th}}$ sample reports a random number $n_C^{MC}(t_i|r, t_0, j)$ of confirmed new cases. We thus average over all the samples and obtain an averaged number of newly confirmed cases on a date $t_i$, $n_C^{MC}(t_i|r, t_0) \coloneqq \sum_{j=1}^{8192} n_C^{MC}(t_i|r, t_0, j)$, and compare it to the actual data $n_C^{Data}(t_i)$. We quantify the quality of the fit by computing the sum of the squared residuals:

$$\varepsilon^2(r, t_0) \coloneqq \sum_{t_i=18}^{26} \left[n_C^{MC}(t_i|r, t_0) - n_C^{Data}(t_i)\right]^2. \tag{7}$$

A $100 \times 100$ grid-based parameter scan is performed to identify the parameters in the region $0.22 < r < 0.42$ and $-20 \leq t_0 \leq -5$ for identifying the best-fit parameters:

$$r^*, t_0^* \coloneqq \operatorname{argmin}_{\{r, t_0\}} \varepsilon^2(r, t_0). \tag{8}$$

As for uncertainty quantification, we formulate the logarithm of the likelihood $\mathcal{L}$ of a parameter set $(r, t_0)$ as

$$\log \mathcal{L}(\alpha, \tau) \coloneqq -n \frac{\varepsilon^2(r, t_0)}{\varepsilon^2(r^*, t_0^*)}. \tag{9}$$

Here, $n = 9$ is the number of data points we use to fit the model. The assumption we make to formulate the above likelihood is that (1) the data (number reported new cases on date $t_i$) is normally distributed with a mean which equals to the Monte Carlo mean reported new cases in our model, and (2) the variance of the noise is identically and $t_i$-independently distributed, and the variance is equal to the mean squared residuals of the best-fit model.

We can then formulate a likelihood ratio test, which quantifies how likely a set of parameters $(r, t_0)$ is in comparison to the best-fit parameters $(r^*, t_0^*)$:

$$\mathbb{P}\{r, t_0 \mid Data\} \sim \exp\left[-n\left(1 - \frac{\varepsilon^2(r, t_0)}{\varepsilon^2(r^*, t_0^*)}\right)\right]. \tag{10}$$

In Bayesian inference, what we computed is essentially the joint posterior distribution of the model parameters $(r, t_0)$, provided a uniform prior distribution on the region of our interests. We present this joint distribution in Supplementary Fig. S4. Finally, because the joint posterior is narrowly distributed, we can numerically compute the marginalized posterior,

$$\begin{aligned}\mathbb{P}\{r \mid Data\} &\sim \int \mathbb{P}\{r, t_0 \mid Data\}\, \mathrm{d}t_0,\\ \mathbb{P}\{t_0 \mid Data\} &\sim \int \mathbb{P}\{r, t_0 \mid Data\}\, \mathrm{d}r,\end{aligned} \tag{11}$$

which is reported in Fig. 2D-F and used to calculate the bounds of centered $95\%$ probability mass to estimate the confidence interval of the growth rate $r$.

**Calculation of *$R_0$* from exponential growth rate**

Assuming gamma distributions for the latent and infectious periods, Wearing et al. (4) have shown that the value of $R_0$ can be calculated from estimated exponential growth rate, *r*, of an outbreak as:

$$R_0 = \frac{r\left(\frac{r}{\sigma m}+1\right)^m}{\gamma\left[1-\left(\frac{r}{\gamma n}+1\right)^{-n}\right]}, \tag{12}$$

where $1/\sigma$ and $1/\gamma$ are the mean latent and infectious periods, respectively, and $m$ and $n$ are the shape parameters for the gamma distributions for the mean latent and infectious periods, respectively.

To quantify the uncertainty of $R_0$, we assume the parameters $(r, \sigma, \gamma, m, n)$ are mutually independent and we generate random samples to compute the resulting $R_0$. Specifically, we generate the samples according to

1. $r \sim \mathbb{P}\{\alpha \mid Data\}$, i.e., we resample the posterior distribution from Eq. (11),
2. $m = 4.5$,
3. $n \sim \mathrm{Unif}(1,6)$,
4. $1/\gamma \sim \mathrm{Unif}(2,8)$ in the first scenario, and $\mathrm{Unif}(4,10)$ in the second scenario.
5. $\sigma \sim \mathcal{N}(\mu = 1/4.2,\ \sigma = 0.0245)$ in the first scenario, and $\mathcal{N}(\mu = 1/2.2,\ \sigma = 0.0468)$.

we generate $10^5$ parameters and compute their respective $R_0$ using Eq. (12). The resulting evaluation were binned into 40 bins to generate histograms. We used the 97.5% and 2.5% percentile of the generate data to quantify the 95% confidence interval.

**Calculation of the impact of intervention strategies**

Using a susceptible–exposed (noninfectious)– infectious–recovered (SEIR) type compartmental model, Lipsitch et al. (*5*) evaluated the impact of quarantine of symptomatic cases to prevent further transmission and quarantine and close observation of asymptomatic contacts of cases so that they may be isolated when they show possible signs of the disease. Assuming that only symptomatic individuals transmit the pathogen, they showed that the reproductive number after the intervention, $R_{int}$, can be expressed as:

$$R_{int} = \frac{R(1-q)D_{int}}{D}, \tag{13}$$

where $R$ is the reproductive number before intervention, $q$ is the percentage of infected individuals being quarantined, $D_{int}$ and $D$ are the mean durations of infectious period after intervention and without intervention, respectively.

Here in our model, we adopted this formulation; however, we assumed that a fraction, $f$, of infected individuals are asymptomatic and can transmit. In this case, quarantine of symptomatic individuals only reduces the contribution of these individuals towards the reproductive number. Thus, we can calculate the reproductive number under quarantine, $R_q$, as:

$$R_q = fR + (1-f)R_{int} = R\left(f + (1-f)(1-q)\frac{D_{int}}{D}\right). \quad (14)$$

We also considered another form of control measure, i.e. the population-level control measure that reduces overall number of daily contacts in the population by $\varepsilon$. These measures include closing down of transportation systems, work and/or school closure, etc. Since *R* depends linearly on the number of daily contacts, we calculate the combined impact of the individual-based quarantine and the population level control measure as:

$$R_{combine} = (1-\varepsilon)[fR + (1-f)R_{int}] = (1-\varepsilon)R\left(f + (1-f)(1-q)\frac{D_{int}}{D}\right). \quad (15)$$

In our calculations, we assumed that the mean duration of infectious period of 2019-nCoV to be 5 days, i.e. $D$=5 days and that $D_{int} = 2$ days. We set the value of $R$ to be the maximum likelihood estimate of $R_0$. Then the impact of the two types of interventions are calculated.

**Fitting the number of new cases in and out of Hubei**

To infer the growth rate of the number of new cases, we used linear regression over the log-transformed case counts. We used the day in January 2020 as an independent variable. For this specific analysis, we avoided using case frequencies $< 10$ because infection dynamics may have been dominated by stochasticity. For cases inside Hubei, we used the number of cases reported between Jan. 16 and Feb. 4. For cases outside of Hubei, we used the number of cases reported between Jan 20. and Feb. 4. To assess whether a different growth rate was observed after Jan 25 outside of Hubei, we evaluated the significance of the interaction term between variable day and the index variable for dates Jan 25 and beyond; the results are presented in Fig. 3C. All regressions and confidence interval estimates were obtained through software R.

**Supplementary references:**

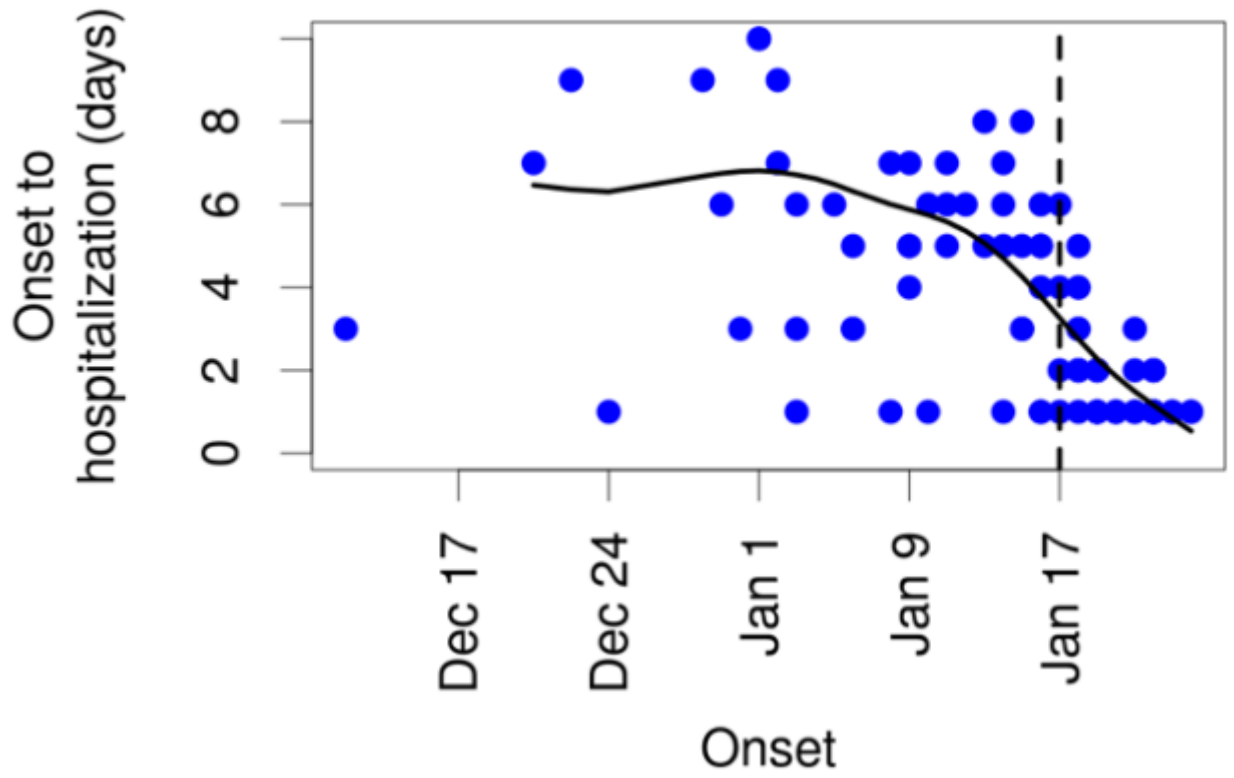

**Figure S1.** The duration from symptom onset to hospitalization decreases over time during the outbreak.

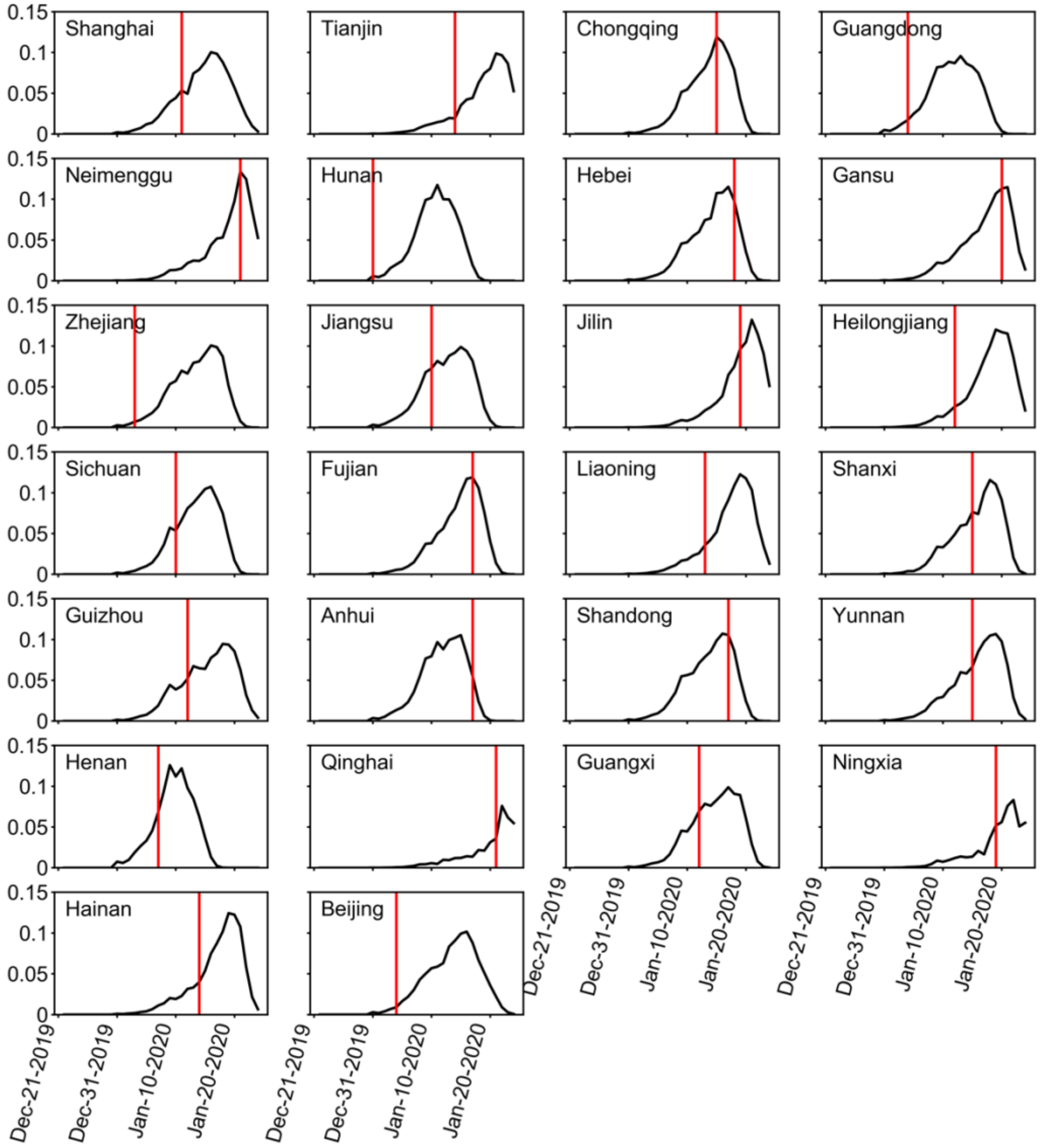

**Figure S2.** Predictions of the 'first arrival' model using best-fit parameters agree well with data. Probability densities of times of first arrival of infected cases in each province based on our maximum likelihood estimate (curves) and documented times of first arrival of infected individuals in our case report dataset (lines).

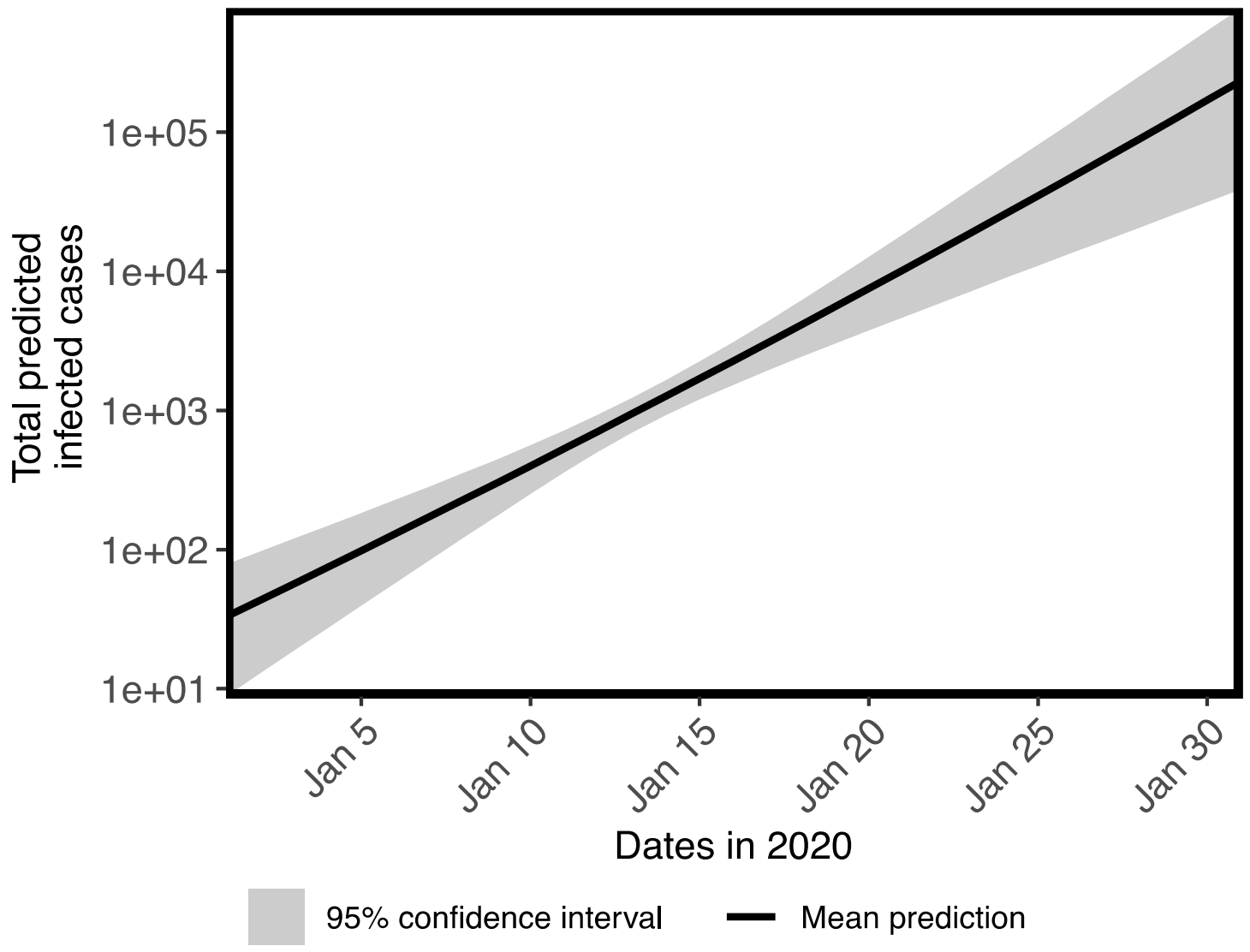

**Figure S3.** Projections of numbers of infected individuals in Wuhan between January 1 and 30, 2020 using the likelihood profile of parameter values in the 'first arrival' approach. Projections after the lock-down of Wuhan on January 23 were hypothetical scenarios assuming no control measures are implemented.

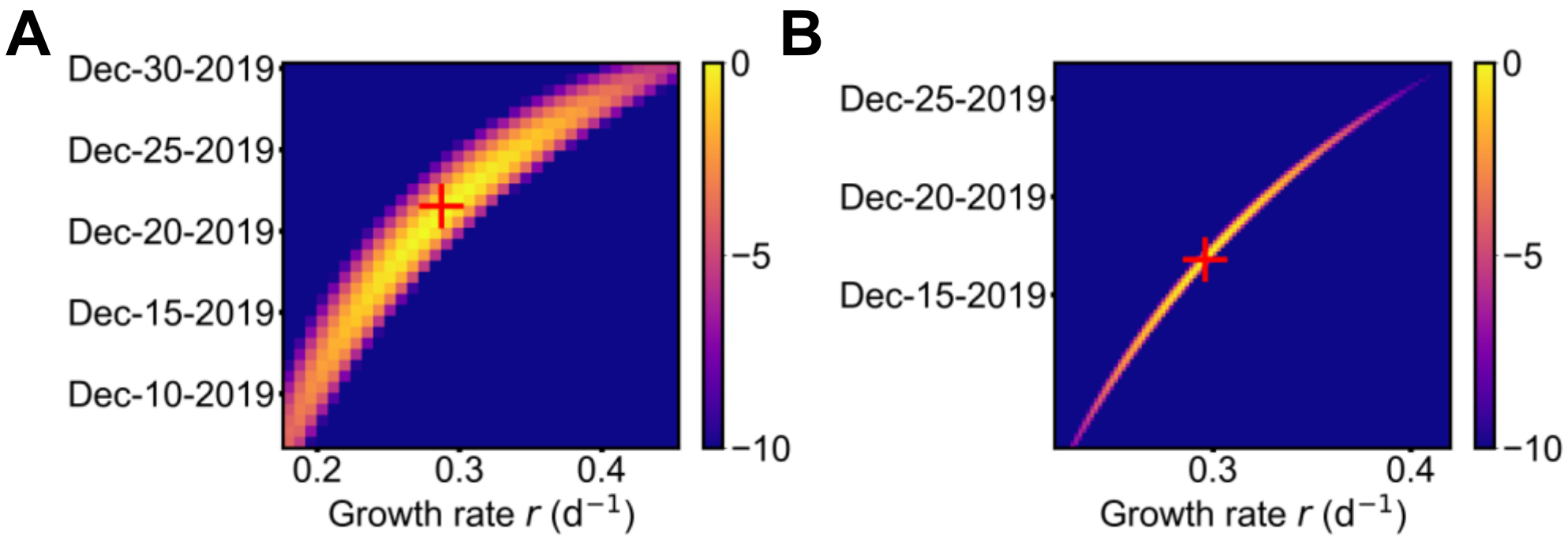

**Figure S4.** Log-likelihood profiles of the estimated exponential growth rate of the outbreak, r (x-axis) and the date of exponential growth initiation (y-axis) from the 'first arrival' model (A) and the 'case count' model (B).

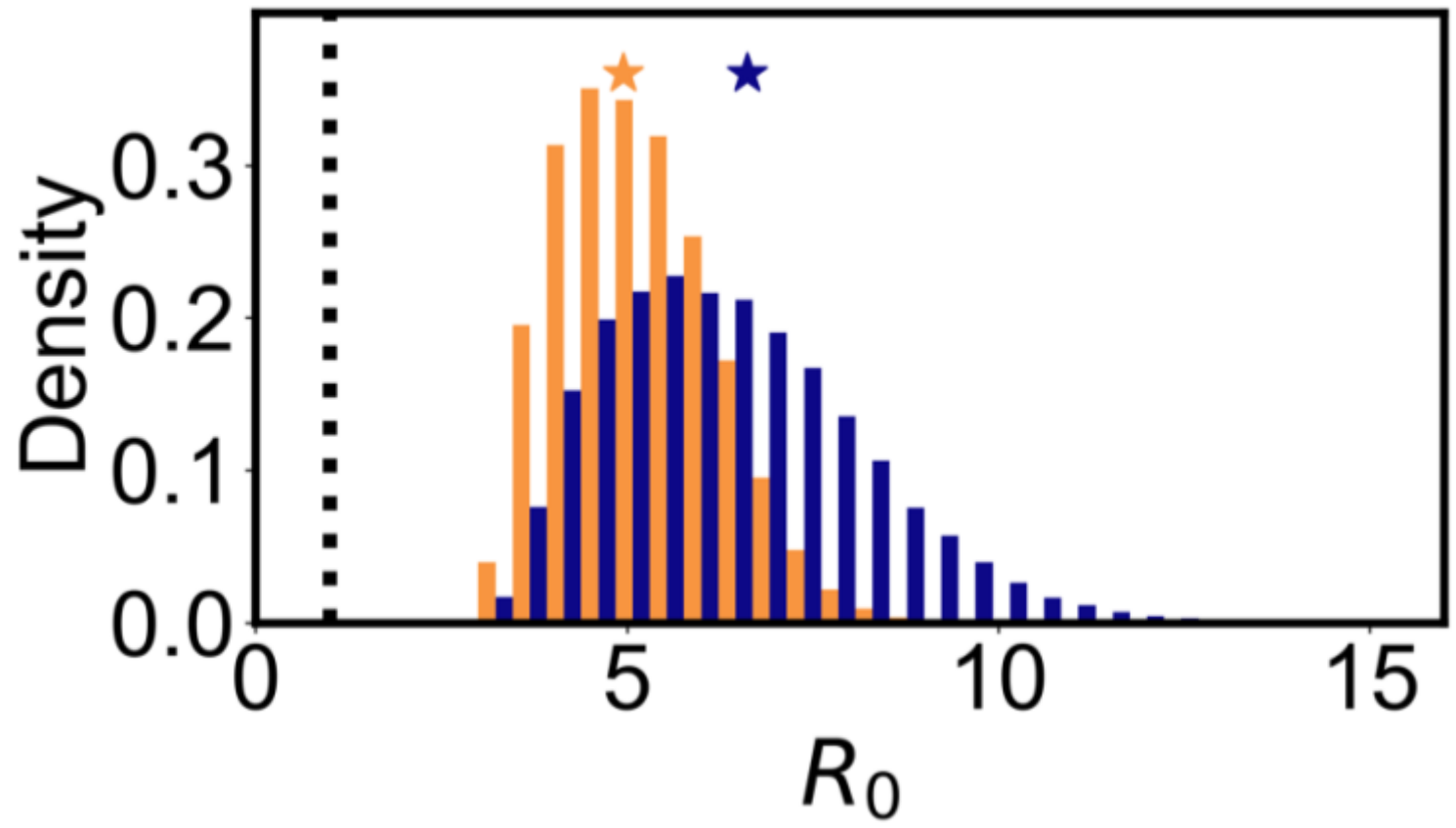

**Figure S5.** Histogram of the basic reproductive number, $R_0$, using the 'case count' model assuming individuals become infectious at symptom onset (blue) or 2 days before symptom onset (orange). The mean estimates are $R_0$=6.6 (blue star) with a CI between 4.0 and 10.5 and $R_0$=4.9 (orange star) with a CI between 3.3 to 7.2. The dashed line denotes $R_0$=1.

## Table S1. Case reports of 2019-nCoV infected individuals

| Province/ city/country | Age | Gender | First day of exposure in Wuhan (if applicable) | Last day of exposure in Wuhan (if applicable) | Departure from Wuhan (if applicable) | Onset date | Hospitalization date | Confirmation date | Date of discharge or death (death cases are commented) | Comment |
|---|---|---|---|---|---|---|---|---|---|---|
| Anhui | | | | | 1/17/20 | | | | | First confirmed case |
| Beijing | | Male | 1/7/20 | 1/08/20 | 1/8/20 | 1/13/20 | | 1/20/20 | 1/25/20 | |
| Beijing | | | 1/9/20 | 1/11/20 | 1/11/20 | 1/14/20 | | 1/20/20 | | |
| Beijing | | Female | | | | 1/13/20 | | 1/20/20 | | |
| Beijing | 45 | Male | 1/11/20 | 1/14/20 | 1/14/20 | 1/19/20 | 1/21/20 | 1/22/20 | | |
| Beijing | 42 | Male | 1/18/20 | 1/18/20 | 1/19/20 | 1/20/20 | 1/20/20 | 1/22/20 | | |
| Beijing | 33 | Female | | | | 1/18/20 | 1/20/20 | 1/22/20 | | |
| Beijing | 33 | Female | | | 1/18/20 | | 1/20/20 | 1/22/20 | | |
| Beijing | | Female | | | 1/8/20 | 1/8/20 | | | 1/24/20 | |
| Beijing | 37 | Male | 1/10/20 | 1/11/20 | 1/11/20 | 1/14/20 | 1/20/20 | 1/21/20 | | |
| Beijing | 39 | Male | 1/3/20 | 1/4/20 | 1/4/20 | 1/9/20 | 1/14/20 | 1/21/20 | | |
| Beijing | 56 | Male | 1/8/20 | 1/16/20 | 1/16/20 | 1/16/20 | 1/20/20 | 1/21/20 | | |
| Beijing | 18 | Female | 1/12/20 | 1/17/20 | 1/17/20 | 1/19/20 | 1/20/20 | 1/21/20 | | |
| Beijing | 32 | Female | 1/13/20 | 1/14/20 | 1/17/20 | 1/15/20 | 1/20/20 | 1/21/20 | | |
| Beijing | | Male | 1/14/20 | 1/14/20 | | 1/18/20 | 1/20/20 | 1/25/20 | | |
| Beijing | 50 | Male | | | | 1/13/20 | 1/21/20 | 1/23/20 | | |
| Beijing | 35 | Male | | | | 1/19/20 | 1/21/20 | 1/23/20 | | |
| Beijing | 36 | Male | | | | 1/19/20 | 1/21/20 | 1/23/20 | | |
| Beijing | 37 | Male | | | | 1/17/20 | 1/19/20 | 1/23/20 | | |
| Beijing | 23 | Female | | | | 1/14/20 | 1/21/20 | 1/23/20 | | |
| Beijing | 33 | Female | | | | 1/17/20 | 1/17/20 | 1/23/20 | | |
| Beijing | 49 | Male | | | | 1/18/20 | 1/21/20 | 1/23/20 | | |
| Beijing | 55 | Female | | | | 1/18/20 | 1/21/20 | 1/23/20 | | |
| Beijing | 44 | Male | | | | 1/18/20 | 1/22/20 | 1/23/20 | | |
| Beijing | 65 | Male | | | | 1/22/20 | 1/22/20 | 1/23/20 | | |
| Beijing | 21 | Male | | | | 1/19/20 | 1/19/20 | 1/23/20 | | |
| Beijing | 41 | Male | | | | 1/20/20 | 1/21/20 | 1/23/20 | | |
| Beijing | 40 | Female | | | | 1/17/20 | 1/23/20 | 1/24/20 | | |
| Beijing | 35 | Male | | | | 1/14/20 | 1/15/20 | 1/24/20 | | |
| Beijing | 29 | Female | | | | 1/22/20 | 1/22/20 | 1/24/20 | | |
| Beijing | 42 | Male | | | | 1/22/20 | 1/23/20 | 1/24/20 | | |
| Beijing | 55 | Male | | | | 1/18/20 | 1/22/20 | 1/24/20 | | |
| Beijing | 50 | Male | | | | 1/15/20 | 1/23/20 | 1/24/20 | | |
| Beijing | 55 | Female | | | | 1/18/20 | 1/23/20 | 1/24/20 | | |
| Beijing | 36 | Female | | | | 1/22/20 | 1/22/20 | 1/25/20 | | |
| Beijing | 44 | Female | | | | 1/10/20 | 1/23/20 | 1/25/20 | | |
| Beijing | 45 | Male | | | | 1/12/20 | 1/18/20 | 1/25/20 | | |
| Beijing | 42 | Male | | | | 1/22/20 | 1/23/20 | 1/25/20 | | |
| Beijing | 37 | Male | | | | 1/18/20 | 1/20/20 | 1/25/20 | | |
| Beijing | 31 | Female | | | | 1/21/20 | 1/24/20 | 1/25/20 | | |
| Beijing | 61 | Female | | | | 1/23/20 | 1/23/20 | 1/25/20 | | |
| Beijing | 47 | Male | | | | 1/22/20 | 1/24/20 | 1/25/20 | | |
| Beijing | 63 | Male | | | | 1/21/20 | 1/23/20 | 1/25/20 | | |
| Beijing | 56 | Female | | | | 1/24/20 | 1/24/20 | 1/25/20 | | |
| Beijing | 43 | Male | | | | 1/23/20 | 1/24/20 | 1/25/20 | | |
| Beijing | 17 | Female | | | | 1/22/20 | 1/24/20 | 1/25/20 | | |
| Beijing | 33 | Female | | | | 1/18/20 | 1/18/20 | 1/25/20 | | |
| Beijing | 78 | Female | | | | 1/24/20 | 1/24/20 | 1/25/20 | | |
| Beijing | 32 | Male | | | | 1/22/20 | 1/24/20 | 1/25/20 | | |
| Chongqing | 44 | Female | | | 1/15/20 | | | 1/21/20 | | First confirmed case |
| Fujian | 70 | Male | | | 1/17/20 | | 1/20/20 | 1/22/20 | | |
| Gansu | 24 | Male | | | 1/20/20 | 1/16/20 | 1/22/20 | 1/23/20 | | First confirmed case |
| Guangdong | 35 | Male | | | 1/15/20 | 1/9/20 | 1/16/20 | | 1/23/20 | |
| Guangdong | 66 | Male | 12/29/19 | 1/2/20 | 1/4/20 | 1/3/20 | 1/4/20 | 1/19/20 | | First confirmed case |
| Guangdong | 10 | Male | 12/29/19 | 12/31/19 | 1/4/20 | 1/1/20 | 1/11/20 | | 1/23/20 | |
| Guangdong | 65 | Female | 12/29/19 | 12/29/19 | 1/4/20 | 1/3/20 | 1/9/20 | | | |
| Guangdong | 37 | Female | 12/29/19 | 12/29/19 | 1/4/20 | 1/2/20 | 1/11/20 | | | |
| Guangdong | 36 | Male | 12/29/19 | 12/31/19 | | 1/1/20 | 1/11/20 | | | |
| Guangdong | 63 | Female | 1/4/20 | 1/7/20 | | 1/8/20 | 1/15/20 | | | |
| Guangxi | 66 | Female | | | 1/12/20 | 1/16/20 | 1/17/20 | 1/22/20 | | First confirmed case |
| Guangxi | 46 | Male | | | | 1/20/20 | 1/21/20 | 1/22/20 | | |
| Guangxi | | | | | 1/20/20 | 1/21/20 | 1/21/20 | 1/22/20 | | |
| Guangxi | 49 | Female | | | 1/15/20 | 1/21/20 | 1/22/20 | 1/25/20 | | |
| Guangxi | 2 | Female | | | 1/21/20 | 1/22/20 | 1/23/20 | 1/25/20 | | |
| Guizhou | 51 | Male | | | 1/12/20 | | 1/14/20 | 1/22/20 | | First confirmed case |
| Hainan | | | | | 1/14/20 | | | | | First confirmed case |
| Hebei | 72 | Male | | | 1/18/20 | 1/19/20 | 1/19/20 | 1/22/20 | | First confirmed case |
| Heilongjiar | 69 | Male | | | 1/12/20 | 1/12/20 | | 1/23/20 | | First confirmed case |
| Henan | 66 | Male | | | 1/7/20 | 12/29/19 | 1/7/20 | 1/21/20 | | First confirmed case |
| Henan | 45 | Male | 1/10/20 | 1/22/20 | | 1/23/20 | 1/23/20 | 1/26/20 | | |
| Henan | 47 | Female | 1/10/20 | 1/13/20 | | 1/14/20 | 1/24/20 | 1/26/20 | | |
| Henan | 48 | Female | 1/10/20 | 1/24/20 | | 1/25/20 | 1/25/20 | 1/26/20 | | |
| Hubei | 23 | Male | | | | 12/24/19 | 12/25/19 | | 1/15/20 | |
| Hunan | 57 | Female | | | | | 1/16/20 | 1/21/20 | | First confirmed case |
| Hunan | 35 | Female | 12/20/19 | 12/31/19 | 12/31/19 | | | 1/22/20 | | |
| Hunan | 40 | Female | 12/20/19 | 12/31/19 | 12/31/19 | | | 1/22/20 | | |
| Hunan | 40 | Male | | | | 1/5/20 | 1/20/20 | 1/23/20 | | |
| Hunan | 45 | Male | | | | 1/16/20 | 1/21/20 | 1/23/20 | | |
| Hunan | 66 | Female | | | | 1/17/20 | 1/21/20 | 1/23/20 | | |
| Hunan | 59 | Male | | | | 1/16/20 | 1/21/20 | 1/23/20 | | |
| Hunan | 23 | Female | | | | 1/16/20 | 1/21/20 | 1/23/20 | | |
| Japan | 30 | Male | | | 1/6/20 | | | | | First confirmed case |
| Jiangsu | 37 | Male | | | 1/10/20 | 1/10/20 | 1/10/20 | 1/22/20 | 1/24/20 | First confirmed case |
| Jilin | 42 | Female | | | 1/19/20 | | 1/19/20 | 1/22/20 | | First confirmed case |
| Liaoning | 33 | Male | | | 1/17/20 | 1/11/20 | 1/17/20 | 1/22/20 | | First confirmed case |
| Liaoning | 40 | Male | | | 1/13/20 | 1/14/20 | 1/19/20 | 1/22/20 | | |
| Liaoning | 50 | Female | | | 1/15/20 | 1/16/20 | 1/16/20 | 1/23/20 | | |
| Liaoning | 48 | Female | 1/25/20 | 1/25/20 | | | 1/29/20 | 1/30/20 | | Asymptomatic at least until 1/30 |
| Macau | 66 | Male | | | 1/22/20 | 1/22/20 | 1/22/20 | 1/23/20 | | |
| Mexico | 57 | Male | 12/25/19 | 1/10/20 | 1/10/20 | | | | | First suspected case |
| Neimenggu | 30 | Male | | | 1/21/20 | | 1/21/20 | 1/24/20 | | |
| Ningxia | 29 | Male | | | 1/19/20 | | | | | First confirmed case |
| Qinghai | 27 | Male | | | 1/21/20 | | 1/23/20 | 1/24/20 | | First confirmed case |
| Shandong | 37 | Male | | | | | 1/17/20 | 1/22/20 | | First confirmed case |
| Shanghai | 56 | Female | | | 1/12/20 | | 1/15/20 | 1/20/20 | 1/23/20 | First confirmed case, No sympto |
| Shanghai | 35 | Male | 1/8/20 | 1/11/20 | 1/11/20 | 1/11/20 | 1/16/20 | 1/21/20 | | |
| Shannxi | 49 | Male | | | 1/19/20 | 1/19/20 | 1/21/20 | 1/24/20 | | |
| Shannxi | 23 | Male | | | 1/22/20 | | 1/22/20 | 1/24/20 | | |
| Shanxi | | Male | 1/12/20 | 1/15/20 | 1/15/20 | 1/19/20 | 1/20/20 | 1/22/20 | | First confirmed case |
| Sichuan | 50 | Male | | | 1/13/20 | | 1/18/20 | 1/23/20 | | |
| Sichuan | 48 | Male | | | 1/10/20 | | 1/18/20 | 1/23/20 | | |
| Sichuan | 36 | Male | | | 1/17/20 | 1/18/20 | 1/20/20 | 1/23/20 | | |
| Sichuan | 34 | Male | | | | | 1/11/20 | 1/21/20 | | First confirmed case |
| Sichuan | 57 | Male | | | 1/15/20 | | 1/16/20 | 1/22/20 | | |
| Sichuan | 28 | Female | | | 1/17/20 | | 1/19/20 | 1/22/20 | | |
| Sichuan | 37 | Male | | | 1/18/20 | | 1/20/20 | 1/22/20 | | |
| Sichuan | 19 | Male | | | 1/13/20 | | 1/20/20 | 1/22/20 | | |
| South Kore | 35 | Female | | | 1/19/20 | | | 1/20/20 | | First confirmed case |
| Tailand | 33 | Female | | | 1/21/20 | | | | | |
| Tailand | 61 | Female | | | | | | 1/13/20 | 1/15/20 | Firstconfirmed case |
| Taiwan | | Female | 1/22/20 | 01/22/20 | | 01/25/20 | | | | |
| Taiwan | 50 | Female | 1/13/20 | 01/15/20 | | 01/22/20 | | | | |
| Taiwan | | | 1/20/20 | 01/20/20 | | 01/25/20 | | | | |
| Tianjin | 60 | Female | | | 1/19/20 | | 1/19/20 | 1/21/20 | | |
| Tianjin | 58 | Male | | | 1/14/20 | | 1/14/20 | 1/21/20 | | First confirmed case |
| Wuhan | 69 | Female | | | | | 1/14/20 | | 1/22/20 | Death |
| Wuhan | 36 | Male | | | | 1/6/20 | 1/9/20 | | 1/23/20 | Death |
| Wuhan | 73 | Male | | | | 12/30/19 | 1/5/20 | | 1/22/20 | Death |
| Wuhan | 70 | Female | | | | 1/15/20 | 1/18/20 | | 1/23/20 | Death |
| Wuhan | 81 | Male | | | | 1/9/20 | 1/13/20 | | 1/21/20 | Death |
| Wuhan | 65 | Female | | | | | 1/13/20 | 1/23/20 | 1/23/20 | Death |
| Wuhan | 61 | Male | | | | 12/20/19 | 12/27/19 | | 1/9/20 | Death |
| Wuhan | 69 | Male | | | | 12/31/19 | 1/3/20 | | 1/15/20 | Death |
| Wuhan | 89 | Male | | | | 1/8/20 | 1/9/20 | | 1/18/20 | Death |
| Wuhan | 89 | Male | | | | 1/13/20 | 1/18/20 | | 1/19/20 | Death |
| Wuhan | 66 | Male | | | | 1/10/20 | 1/16/20 | | 1/20/20 | Death |
| Wuhan | 75 | Male | | | | 1/6/20 | 1/11/20 | | 1/20/20 | Death |
| Wuhan | 48 | Female | | | | 12/10/19 | 12/13/19 | | 1/20/20 | Death |
| Wuhan | 82 | Male | | | | 1/9/20 | 1/14/20 | | 1/21/20 | Death |
| Wuhan | 66 | Male | | | | 12/22/19 | 12/31/19 | | 1/21/20 | Death |
| Wuhan | 81 | Male | | | | 1/15/20 | 1/18/20 | | 1/22/20 | Death |
| Wuhan | 82 | Female | | | | 1/3/20 | 1/6/20 | | 1/22/20 | Death |
| Wuhan | 65 | Male | | | | 1/5/20 | 1/11/20 | | 1/21/20 | Death |
| Wuhan | 80 | Female | | | | 1/11/20 | 1/18/20 | | 1/22/20 | Death |

| | | | | | | | | | | | |
|---|---|---|---|---|---|---|---|---|---|---|---|
| Wuhan | 53 | Male | | | | | 1/5/20 | | 1/21/20 | Death |
| Wuhan | 86 | Male | | | | 1/2/20 | 1/9/20 | | 1/21/20 | Death |
| Wuhan | 70 | Female | | | | | 1/13/20 | | 1/21/20 | Death |
| Wuhan | 84 | Male | | | | 1/6/20 | 1/9/20 | | 1/22/20 | Death |
| Yunnan | 51 | Male | | | 1/15/20 | | 1/16/20 | | | First confirmed case |
| Zhejiang | 46 | Male | | | 1/3/20 | 1/4/20 | 1/17/20 | 1/21/20 | 1/24/20 | First confirmed case, no symptor |

**Table S2. Migration indices to and from Wuhan from Baidu Huiyan and calculated number of travellers out of the province of Hubei.**

| Date | Emigration Index | Immigration Index | Total pop size out of Wuhan | Fraction to other provinces | Pop exported out of Hubei from Wuhan | Total Pop size in Wuhan |
|---|---|---|---|---|---|---|
| 1/1/2020 | 3.46 | 2.85 | 154038 | 0.2777 | 42776 | 13972843 |
| 1/2/2020 | 3.52 | 3.09 | 156709 | 0.3867 | 60599 | 13953700 |
| 1/3/2020 | 5.52 | 4.22 | 245748 | 0.3276 | 80507 | 13895824 |
| 1/4/2020 | 6.1 | 4.45 | 271570 | 0.3226 | 87608 | 13822367 |
| 1/5/2020 | 5.32 | 5.08 | 236844 | 0.3701 | 87656 | 13811682 |
| 1/6/2020 | 5.6 | 4.31 | 249310 | 0.3767 | 93915 | 13754252 |
| 1/7/2020 | 6.41 | 4.25 | 285371 | 0.3787 | 108070 | 13658089 |
| 1/8/2020 | 7.34 | 4.47 | 326774 | 0.3862 | 126200 | 13530318 |
| 1/9/2020 | 8.14 | 4.81 | 362390 | 0.3848 | 139448 | 13382067 |
| 1/10/2020 | 6.62 | 4.6 | 294720 | 0.3819 | 112554 | 13292138 |
| 1/11/2020 | 7.56 | 4.64 | 336568 | 0.3257 | 109620 | 13162141 |
| 1/12/2020 | 6.22 | 4.37 | 276912 | 0.3362 | 93098 | 13079779 |
| 1/13/2020 | 5.76 | 4.83 | 256433 | 0.361 | 92572 | 13038376 |
| 1/14/2020 | 5.46 | 4.08 | 243077 | 0.352 | 85563 | 12976939 |
| 1/15/2020 | 5.91 | 4.06 | 263111 | 0.338 | 88932 | 12894578 |
| 1/16/2020 | 6 | 4 | 267118 | 0.3425 | 91488 | 12805538 |
| 1/17/2020 | 6.44 | 4.4 | 286706 | 0.3304 | 94728 | 12714718 |
| 1/18/2020 | 7.71 | 4.23 | 343246 | 0.3004 | 103111 | 12559790 |
| 1/19/2020 | 7.41 | 4.15 | 329890 | 0.305 | 100617 | 12414656 |
| 1/20/2020 | 8.31 | 4.18 | 369958 | 0.2933 | 108509 | 12230790 |
| 1/21/2020 | 10.74 | 4.24 | 478141 | 0.2816 | 134644 | 11941412 |
| 1/22/2020 | 11.84 | 2.9 | 527112 | 0.2523 | 132990 | 11543407 |
| 1/23/2020 | 11.14 | 1.75 | 495949 | 0.2343 | 116201 | 11125367 |
| 1/24/2020 | 3.89 | 0.88 | 173181 | 0.2754 | 47694 | 10991363 |
| 1/25/2020 | 1.3 | 0.63 | 57876 | 0.2544 | 14724 | 10961535 |

**Table S3. Estimated number of individuals who traveled from Wuhan to provinces outside of Hubei.**
The number of individuals who traveled before that date was approximated as the average over the first seven days of January.

| Province | Before Jan 1st | Date in January 2020 | | | | | | | | | | | |
|---|---|---|---|---|---|---|---|---|---|---|---|---|---|
| | | 1 | 2 | 3 | 4 | 5 | 6 | 7 | 8 | 9 | 10 | 11 | 12 |
| Shanghai | 2937 | 1602 | 2116 | 2924 | 3096 | 3766 | 3316 | 3710 | 4117 | 4023 | 3478 | 3265 | 2354 |
| Tianjin | 608 | 308 | 392 | 565 | 733 | 758 | 773 | 771 | 980 | 978 | 855 | 740 | 609 |
| Chongqing | 2523 | 1309 | 1802 | 2531 | 2824 | 2700 | 3067 | 3539 | 4150 | 5291 | 4362 | 4039 | 3711 |
| Guangdong | 7116 | 4020 | 5892 | 7348 | 7495 | 8242 | 7754 | 8418 | 9542 | 10038 | 8341 | 7472 | 6286 |
| Neimenggu | 480 | 185 | 329 | 418 | 597 | 474 | 648 | 799 | 980 | 1196 | 914 | 774 | 831 |
| Hunan | 8101 | 4760 | 5861 | 9166 | 8826 | 8171 | 9075 | 10730 | 12319 | 13082 | 11494 | 12083 | 9553 |
| Hebei | 2116 | 1047 | 1536 | 1892 | 2281 | 2392 | 2593 | 3196 | 3954 | 4602 | 3684 | 3332 | 2880 |
| Gansu | 949 | 323 | 721 | 762 | 1059 | 1042 | 1321 | 1570 | 1863 | 2102 | 1533 | 1346 | 1357 |
| Zhejiang | 3844 | 2141 | 3134 | 4030 | 3992 | 4358 | 4338 | 4623 | 5588 | 5690 | 4774 | 4645 | 3544 |
| Jiangsu | 4685 | 2434 | 3573 | 4718 | 5106 | 5305 | 5460 | 6193 | 6993 | 7610 | 6513 | 5991 | 4652 |
| Jilin | 407 | 200 | 282 | 393 | 489 | 450 | 499 | 571 | 817 | 833 | 560 | 539 | 554 |
| Heilongjiang | 650 | 339 | 501 | 639 | 706 | 663 | 748 | 970 | 1111 | 1268 | 914 | 976 | 997 |
| Sichuan | 2789 | 1279 | 2131 | 2433 | 3096 | 3174 | 3391 | 4195 | 4934 | 5979 | 4362 | 4308 | 4264 |
| Fujian | 1961 | 940 | 1614 | 2064 | 2281 | 1989 | 2269 | 2568 | 3300 | 3696 | 2888 | 2928 | 2575 |
| Liaoning | 943 | 508 | 674 | 786 | 1032 | 1018 | 1197 | 1427 | 1405 | 1558 | 1267 | 1212 | 1025 |
| Shanxi | 1314 | 632 | 1034 | 1155 | 1602 | 1208 | 1596 | 2055 | 2712 | 3262 | 2446 | 2255 | 2132 |
| Guizhou | 2237 | 986 | 1551 | 2187 | 2553 | 2416 | 2917 | 3310 | 4248 | 4494 | 3006 | 2558 | 2409 |
| Anhui | 5200 | 2988 | 4043 | 5554 | 5812 | 5447 | 5684 | 6649 | 7908 | 8842 | 7427 | 7606 | 5871 |
| Shandong | 2825 | 1294 | 2116 | 2630 | 3123 | 3008 | 3515 | 4309 | 4738 | 5726 | 4598 | 3770 | 3655 |
| Yunnan | 1643 | 832 | 1254 | 1622 | 1738 | 1705 | 2119 | 2254 | 2843 | 2718 | 2181 | 2188 | 1911 |
| Henan | 10736 | 6362 | 7726 | 11182 | 11949 | 11416 | 12092 | 14269 | 16437 | 19243 | 15414 | 16559 | 13652 |
| Qinghai | 233 | 77 | 172 | 172 | 272 | 237 | 349 | 400 | 523 | 399 | 383 | 236 | 360 |
| Guangxi | 1915 | 801 | 1457 | 1892 | 2281 | 2108 | 2194 | 2854 | 3823 | 4566 | 3448 | 3298 | 3323 |
| Ningxia | 266 | 92 | 219 | 197 | 353 | 332 | 324 | 371 | 556 | 797 | 501 | 471 | 443 |
| Hainan | 1248 | 693 | 1050 | 1032 | 1276 | 1232 | 1521 | 1883 | 1765 | 1921 | 1356 | 1212 | 1274 |
| Beijing | 4083 | 1972 | 2977 | 3883 | 3829 | 5234 | 5260 | 5536 | 5980 | 5400 | 4804 | 3904 | 3351 |

**Table S3. Continued**

| Province | 13 | 14 | 15 | 16 | 17 | 18 | 19 | 20 | 21 | 22 | 23 | 24 | 25 |
|---|---|---|---|---|---|---|---|---|---|---|---|---|---|
| Shanghai | 2872 | 2528 | 2394 | 2457 | 2294 | 2059 | 1847 | 1702 | 1578 | 1476 | 1537 | 918 | 284 |
| Tianjin | 564 | 413 | 579 | 534 | 430 | 481 | 462 | 407 | 430 | 369 | 298 | 156 | 46 |
| Chongqing | 3411 | 3476 | 4052 | 3873 | 3785 | 4359 | 4256 | 4698 | 5977 | 5482 | 4959 | 2234 | 532 |
| Guangdong | 6283 | 5299 | 5078 | 5209 | 5304 | 5561 | 5839 | 6141 | 8081 | 8223 | 7687 | 5022 | 1806 |
| Neimenggu | 718 | 535 | 474 | 561 | 516 | 412 | 462 | 518 | 669 | 633 | 446 | 260 | 98 |
| Hunan | 9668 | 9286 | 9393 | 9696 | 10293 | 12872 | 11315 | 12431 | 16257 | 17078 | 15226 | 5403 | 1366 |
| Hebei | 2898 | 2455 | 3052 | 2858 | 3211 | 3329 | 3299 | 3478 | 3921 | 3742 | 2777 | 1022 | 307 |
| Gansu | 1333 | 1167 | 1079 | 935 | 946 | 961 | 1023 | 1073 | 1339 | 1423 | 1091 | 571 | 168 |
| Zhejiang | 3513 | 3038 | 2999 | 3179 | 3326 | 3913 | 3431 | 3663 | 4255 | 3742 | 3273 | 1559 | 544 |
| Jiangsu | 4565 | 4254 | 4394 | 4407 | 4874 | 5320 | 4585 | 4661 | 5546 | 5429 | 4712 | 2182 | 758 |
| Jilin | 590 | 535 | 500 | 481 | 631 | 584 | 627 | 592 | 717 | 633 | 546 | 312 | 93 |
| Heilongjiang | 872 | 802 | 868 | 908 | 946 | 961 | 1056 | 999 | 1100 | 1054 | 893 | 416 | 150 |
| Sichuan | 3872 | 3695 | 3710 | 3846 | 3670 | 4050 | 4421 | 4476 | 5403 | 5113 | 4116 | 2078 | 677 |
| Fujian | 2616 | 2455 | 2631 | 2858 | 2982 | 3158 | 3332 | 3330 | 4016 | 3901 | 3571 | 1507 | 394 |
| Liaoning | 1077 | 997 | 947 | 1122 | 1118 | 1167 | 1254 | 1258 | 1387 | 1107 | 843 | 329 | 133 |
| Shanxi | 2051 | 1677 | 1710 | 1362 | 1634 | 1819 | 1880 | 2035 | 2486 | 2530 | 2331 | 918 | 226 |
| Guizhou | 2513 | 1945 | 1552 | 1576 | 1462 | 1545 | 1550 | 1665 | 1817 | 1581 | 1438 | 641 | 232 |
| Anhui | 5949 | 5858 | 6657 | 6491 | 7225 | 8272 | 7489 | 8398 | 10854 | 11069 | 9522 | 3412 | 1048 |
| Shandong | 3334 | 3063 | 3157 | 3312 | 3584 | 3879 | 3728 | 3811 | 4781 | 4480 | 3422 | 1593 | 492 |
| Yunnan | 2051 | 1580 | 1447 | 1549 | 1548 | 1545 | 1616 | 1813 | 2199 | 2056 | 1637 | 745 | 313 |
| Henan | 13694 | 14026 | 15866 | 17256 | 18263 | 19222 | 19332 | 23011 | 29549 | 29887 | 26384 | 8624 | 2506 |
| Qinghai | 256 | 243 | 184 | 160 | 115 | 137 | 99 | 111 | 96 | 158 | 99 | 69 | 29 |
| Guangxi | 3077 | 2455 | 2263 | 2164 | 2236 | 2059 | 2507 | 2664 | 2964 | 2899 | 2579 | 1143 | 411 |
| Ningxia | 385 | 267 | 210 | 240 | 143 | 240 | 264 | 222 | 239 | 211 | 99 | 87 | 35 |
| Hainan | 1026 | 948 | 1000 | 1122 | 1089 | 1098 | 1287 | 1369 | 1721 | 1739 | 1587 | 866 | 394 |
| Beijing | 3693 | 3452 | 3420 | 3419 | 3068 | 2506 | 2243 | 2072 | 2247 | 2056 | 1785 | 675 | 243 |